%% file: main.tex
\UseRawInputEncoding
\documentclass[%
reprint,
superscriptaddress,
amsmath,
amssymb,
aps,
prl,
]{revtex4-1}

\usepackage{graphicx}
\usepackage{float}
\usepackage{braket}
\usepackage{subcaption}
\usepackage{comment}
\usepackage{color, ulem}
\usepackage{gensymb}
\usepackage{xcolor}
\usepackage{wrapfig}
\usepackage{amsmath,amsfonts,amsthm,bm,bigints,amssymb}
\usepackage{mathtools}
\usepackage{bbm}
\usepackage{relsize}
\usepackage[version=4]{mhchem}
\usepackage{chemformula}
\usepackage{hhline}
\usepackage{makecell}
\usepackage[dvipsnames]{xcolor}
\usepackage{array}

\input{input.tex}

\usepackage{xr-hyper}
\makeatletter
\newcommand*{\addFileDependency}[1]{
\typeout{(#1)}
\@addtofilelist{#1}
\IfFileExists{#1}{}{\typeout{No file #1.}}
}\makeatother

\newcommand{\myexternaldocument}[1]{%
\externaldocument{#1}%
\addFileDependency{#1.tex}%
\addFileDependency{#1.aux}%
}

\usepackage{hyperref}
\hypersetup{
    colorlinks=false,
    hidelinks=false,
    linkcolor=black,
    filecolor=magenta, 
    citecolor=black,
    urlcolor=black,
    pdfpagemode=FullScreen,
    }

\myexternaldocument{supplementary_info}

\usepackage{natbib}
\begin{document}

\title{Proximity-induced charge density waves in a moir\'e heterobilayer}

\author{Christopher T. S. Cheung}
\affiliation{Departments of Physics and Materials and the Thomas Young Center for Theory and Simulation of Materials, Imperial College London, South Kensington Campus, London SW7 2AZ, UK\\}
\author{Arash A. Mostofi}
\affiliation{Departments of Physics and Materials and the Thomas Young Center for Theory and Simulation of Materials, Imperial College London, South Kensington Campus, London SW7 2AZ, UK\\}
\author{Johannes Lischner}
\affiliation{Departments of Physics and Materials and the Thomas Young Center for Theory and Simulation of Materials, Imperial College London, South Kensington Campus, London SW7 2AZ, UK\\}

\date{\today}

\begin{abstract}
Twisted heterobilayers of two-dimensional materials have emerged as a  platform for studying  emergent phases of matter. In this work, we investigate charge density waves (CDW) in a twisted NbSe$_2$/MoSe$_2$ bilayer using first-principles calculations. We observe CDW formation in both layers, even though MoSe$_2$ does not feature a CDW in its monolayer form. Moreover, we find that the CDW is highly non-uniform with different types of CDWs coexisting in the moir\'e unit cells of both layers. The amplitude of the CDW also varies in the moir\'e unit cell. We assess different mechanisms of CDW formation in the MoSe$_2$ layer and conclude that the dominant one is the steric repulsion between Se atoms across the van der Waals gap. The strength of this effect is highly sensitive to the stacking, explaining the moir\'e variation of the CDW amplitude. Furthermore, the amplitude of induced CDWs are amplified compared to that expected from untwisted bilayers. This is due to the localization of electrons in the low-energy stacking regions in the NbSe$_2$ layer. Our work demonstrates that novel broken-symmetry phases can be induced and enhanced in twisted heterobilayers through proximity effects.
\end{abstract}

\maketitle
\section{Introduction}
Twisted moir\'e bilayers are tunable platforms for studying novel electronic, optical and structural properties~\cite{Cao_Fatemi_Fang_Watanabe_Taniguchi_Kaxiras_Jarillo-Herrero_2018,Cao_Fatemi_Demir_Fang_Tomarken_Luo_Sanchez-Yamagishi_Watanabe_Taniguchi_Kaxiras_2018,Bistritzer_MacDonald_2011,flat_mag_goodwin,attrac_elel_goodwin,local_dirac_el,Chen_Lian_Huang_Su_Rashetnia_Ma_Yan_Blei_Xiang_Taniguchi_et_2022,mose2_opt_expt,Parzefall_Holler_Scheuck_Beer_Lin_Peng_Monserrat_Nagler_Kempf_Korn,Quan_Linhart_Lin_Lee_Zhu_Wang_Hsu_Choi_Embley_Young_2021,phason_maity,chiral_phon_maity,xu2022coexisting,tong2018skyrmions,akram2021skyrmions,xiao2021magnetization}. For example, twisted bilayers of WSe$_2$ and MoSe$_2$ host moir\'e-trapped interlayer excitons~\cite{interlayer_x_transport} and localization of intralayer excitons in twisted WS$_2$/WSe$_2$ bilayers enables realization of the Bose-Hubbard model~\cite{Lian_Meng_Ma_Maity_Yan_Wu_Huang_Chen_Chen_Chen_2023}. While twisted bilayers formed by semiconducting monolayers have been studied extensively, bilayers involving metallic monolayers have received less attention. 

Metallic 2D materials often exhibit symmetry-broken ground states. For example, the metallic transition metal dichalcogenide (TMD) monolayer NbSe$_2$ hosts a charge density wave (CDW) and superconductivity at low temperatures~\cite{unveil_cdw_sc,unveil_cdw_imp,visualize_triangular_cdw}. CDWs in NbSe$_2$ are formed when the Nb atoms displace from their high-symmetry positions to break the translational and threefold rotational symmetries. Different CDW types have been observed in NbSe$_2$ including filled-center CDWs (where Nb atoms move towards Se atoms), hollow-center CDWs (Nb atoms move towards interstitial sites) and hexagonal CDWs (Nb atoms move towards other Nb atoms)~\cite{mcmillan-cdw1,mcmillan-cdw2,mcmillan-cdw3}. As these different CDWs have very similar energies, their relative stability can be controlled by doping and strain~\cite{elas_cdw_guster}. When two NbSe$_2$ monolayers are combined into a twisted homobilayer, the CDWs in each monolayer become highly non-uniform as a consequence of the inhomogeneous strain induced by moir\'e relaxations and different CDW types, including one-dimensional stripe CDWs, coexist in the moir\'e unit cell~\cite{cdw_nbse2_cheung}.

Apart from metallic 2D materials which often feature intrinsic CDWs, it is also possible to induce CDWs in semiconducting monolayers by charge doping or interfacing with other monolayers~\cite{dreher2021proximity,unveil_cdw_imp}. For example, it was demonstrated that CDWs can be induced in the semiconducting monolayer MoS$_2$ by electron doping~\cite{mos2_cdw}. Also, it is possible to induce a CDW in semi-metallic graphene by interfacing it with NbSe$_2$~\cite{gnbse2_sc}. However, the possibility of inducing CDWs in twisted heterobilayers through a proximity effect has not been studied so far.

In this work, we carry out first-principles relaxation of twisted moir\'e bilayers of NbSe$_2$/MoSe$_2$ using density functional theory and use a novel approach for revealing the presence, type and amplitude of inhomogeneous CDWs in the moir\'e unit cell. We find that weak non-uniform CDWs are induced in the MoSe$_2$ layer due to the interaction with the NbSe$_2$. By studying CDWs in untwisted bilayers, we demonstrate that the mechanism underpinning CDW formation in MoSe$_2$ is the steric repulsion between Se atoms across the van der Waals gap between the two layers. Our work demonstrates that proximity effects can induce inhomogeneous symmetry-broken phases in moir\'e bilayers which are absent in the isolated monolayer.

\section{Methodology}
We construct a NbSe$_2$/MoSe$_2$ bilayer with a twist angle of 3.15$^\circ$ (containing 1,986 atoms in the moir\'e unit cell) using the procedure described in Ref.~\cite{moire_def_dft}. The unit cell lattice constant is taken to be that of an untwisted antiparallel NbSe$_2$/MoSe$_2$ bilayer ($\lattconst = 3.42~\Ang$) obtained by performing a first-principles DFT relaxation using the optB88-vdw exchange-correlation functional as implemented in the SIESTA code~\cite{Soler_Artacho_Gale_Garcia_Junquera_Ordejon_Sanchez-Portal_2002,KBM}. This is the lattice constant of the MM stacked untwisted bilayer, which is the most stable stacking. Note that the lattice constants of the AB and XX stacked untwisted bilayers are very similar (both are $3.41~\Ang$) to the MM value. We then perform a first-principles atomic relaxation of the twisted bilayer. The construction of the moir\'e unit cell and the additional computational details are provided in Section~\ref{sec:methods} of the Supplementary Materials.

We analyze the formation of charge density waves (CDWs) in the twisted bilayer using the smeared metal atomic (either Mo or Nb) density. The smeared atomic density at a position $\pos$ is computed by placing gaussian functions on every Mo and Nb atom, and summing over all the gaussian functions within a cutoff radius within a given layer of the bilayer around that point (for computational efficiency)~\cite{cdw_nbse2_cheung}. Mathematically,
\begin{equation}
    \den(\pos)=\frac{1}{2\pi\smearing^2}\sum_{i=1}^{N}\denexp,
    \label{eq:rho_gauss}
\end{equation}
where $N$ is the number of metal atoms in the given layer within the cutoff radius, $\atpos_i$ is the relaxed position of the $i$-th metal atom and $\sigma$ is a smearing parameter. By using a smearing parameter that is comparable to the unit cell lattice constant $\lattconst$, local changes in the Nb--Nb distances give rise to a modulation in the smeared atomic density. This allows us to reveal any CDW formation.  

\newcommand{\cdwhspace}{\hspace{40pt}}
\newcommand{\cdwscale}{0.32\textwidth}
\newcommand{\cdwscaleb}{0.32\textwidth}
\begin{figure*}[htb!]
    \centering
    \begin{subfigure}[h]{\cdwscale}
        \centering
        \caption{\cdwhspace Hollow-center CDW}
        \includegraphics[width=\textwidth]{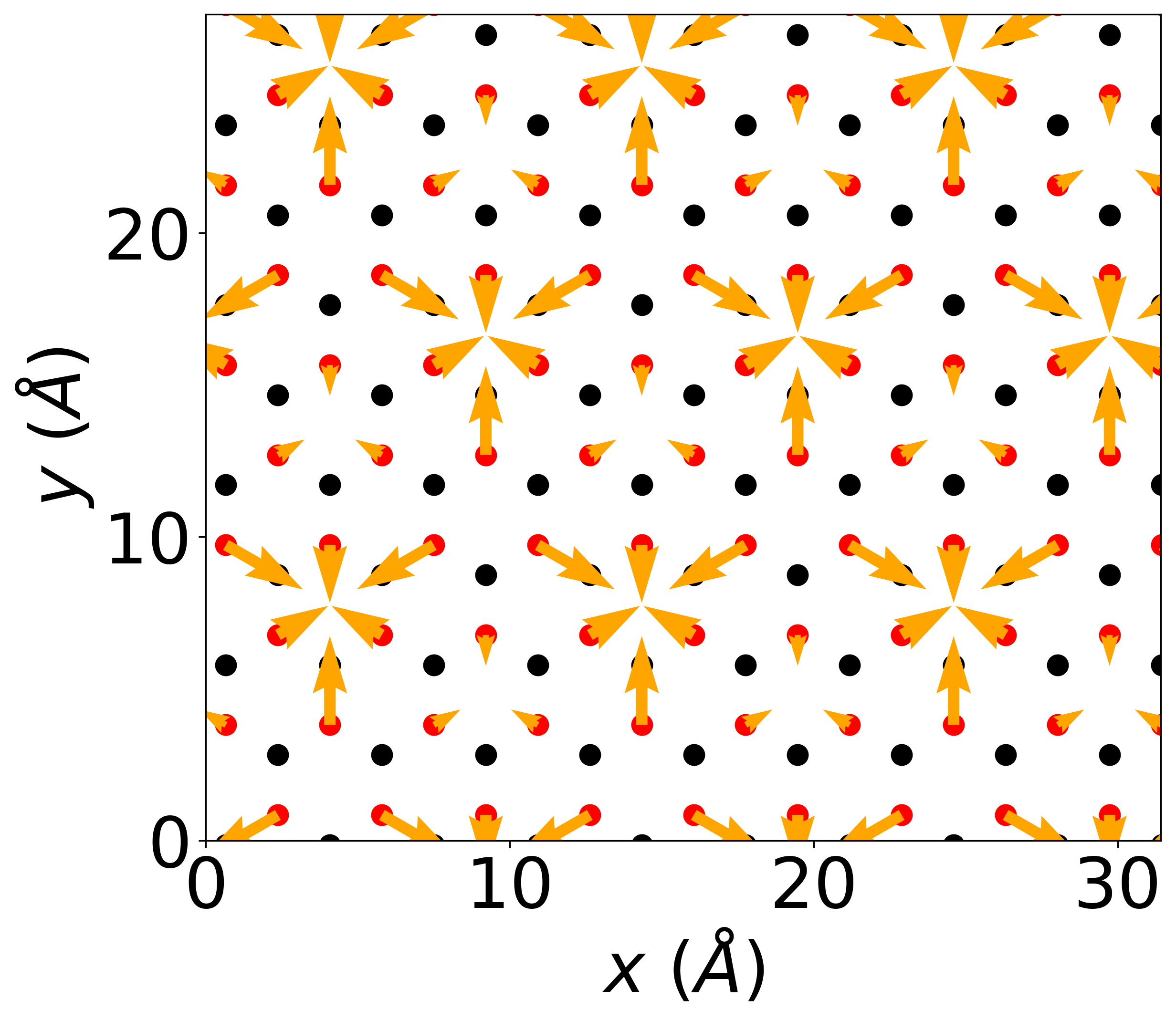}
    \end{subfigure}%
    ~
    \begin{subfigure}[h]{\cdwscale}
        \centering
        \caption{\cdwhspace Filled-center CDW}
        \includegraphics[width=\textwidth]{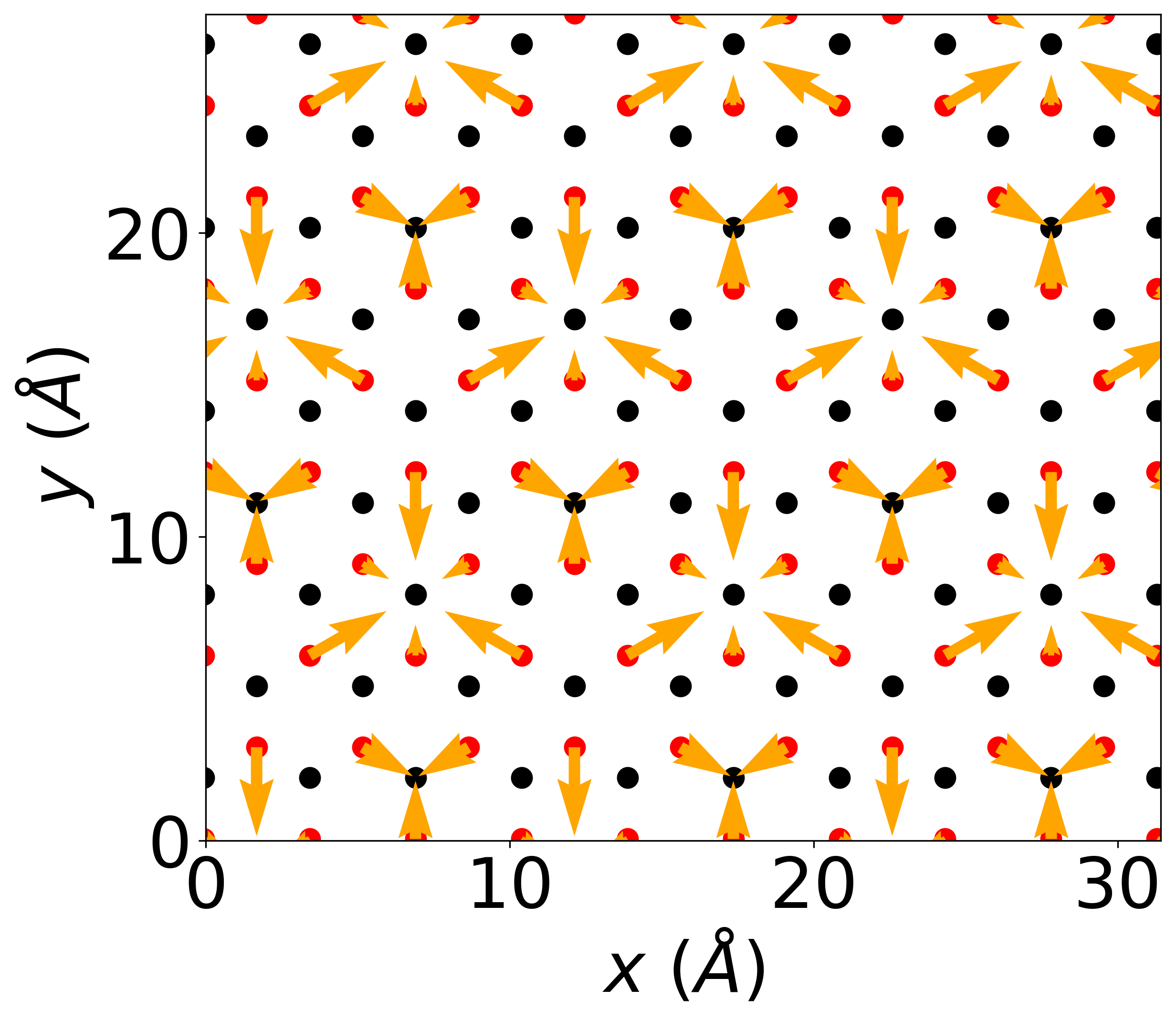}
    \end{subfigure}%
    ~
    \begin{subfigure}[h]{\cdwscale}
        \centering
        \caption{\cdwhspace Hexagonal CDW}
        \includegraphics[width=\textwidth]{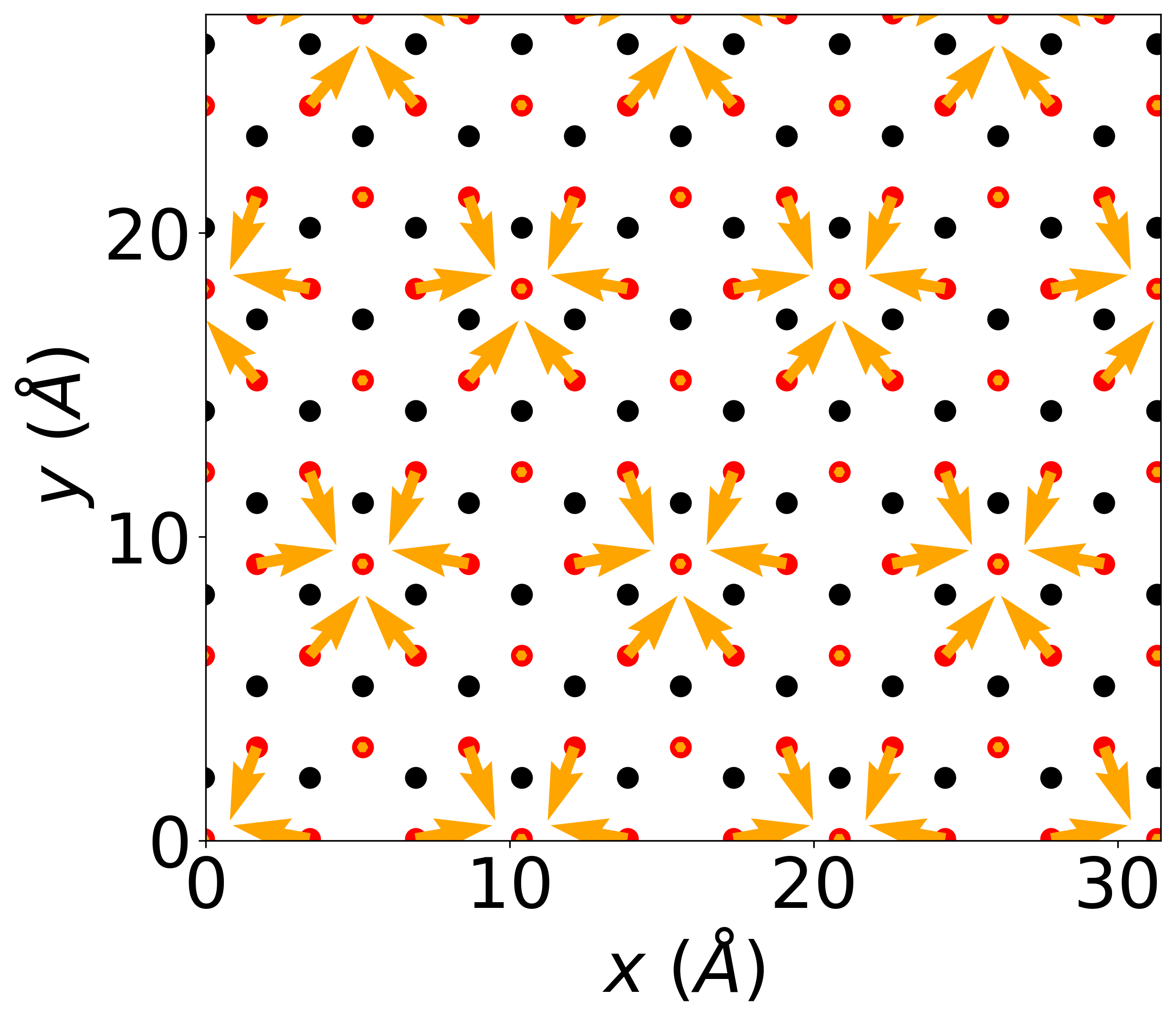}
    \end{subfigure}%
    \par
    \begin{subfigure}[h]{\cdwscaleb}
        \centering
        \caption{}
        \includegraphics[width=\textwidth]{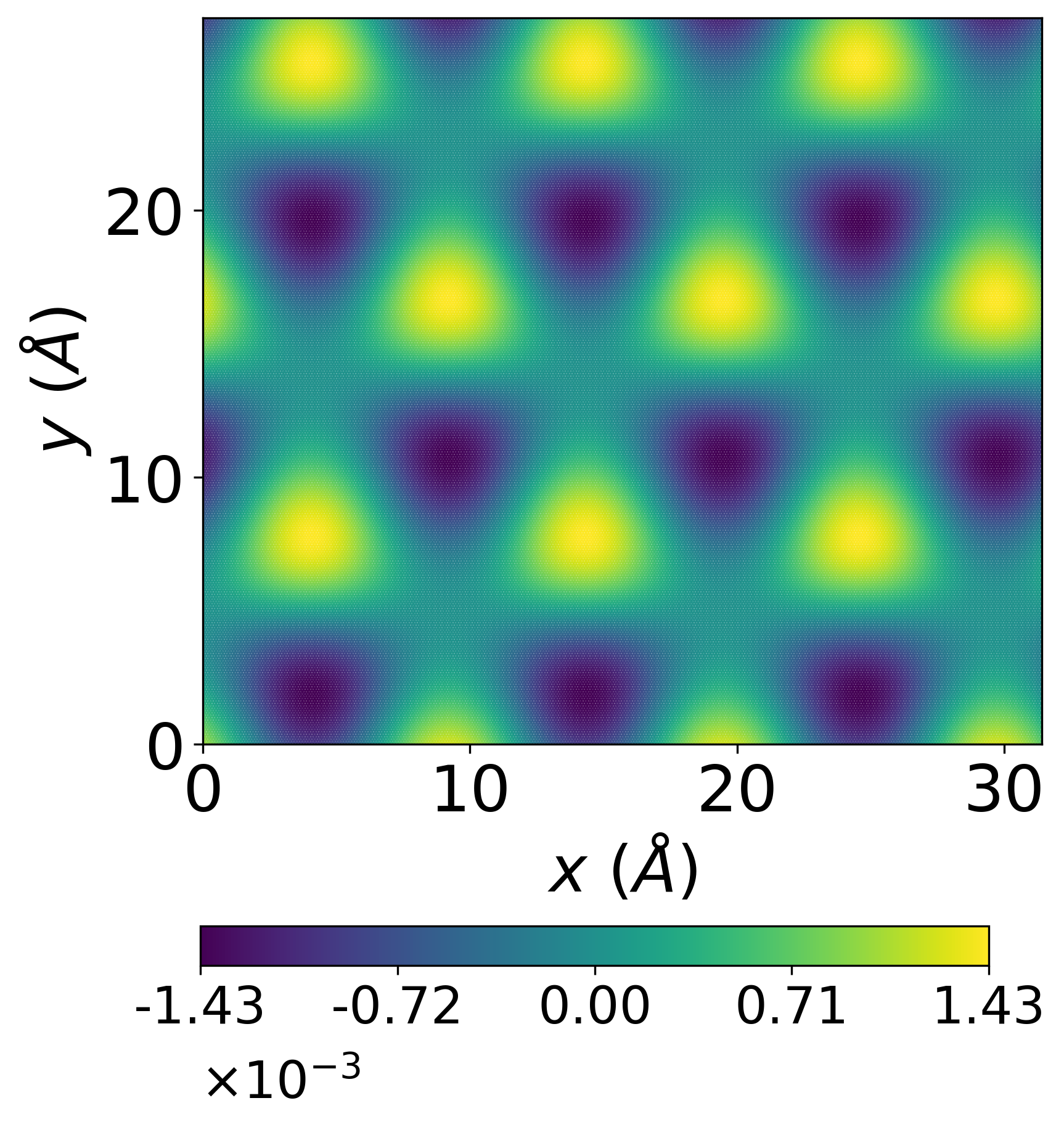}
    \end{subfigure}%
    ~
    \begin{subfigure}[h]{\cdwscaleb}
        \centering
        \caption{}
        \includegraphics[width=\textwidth]{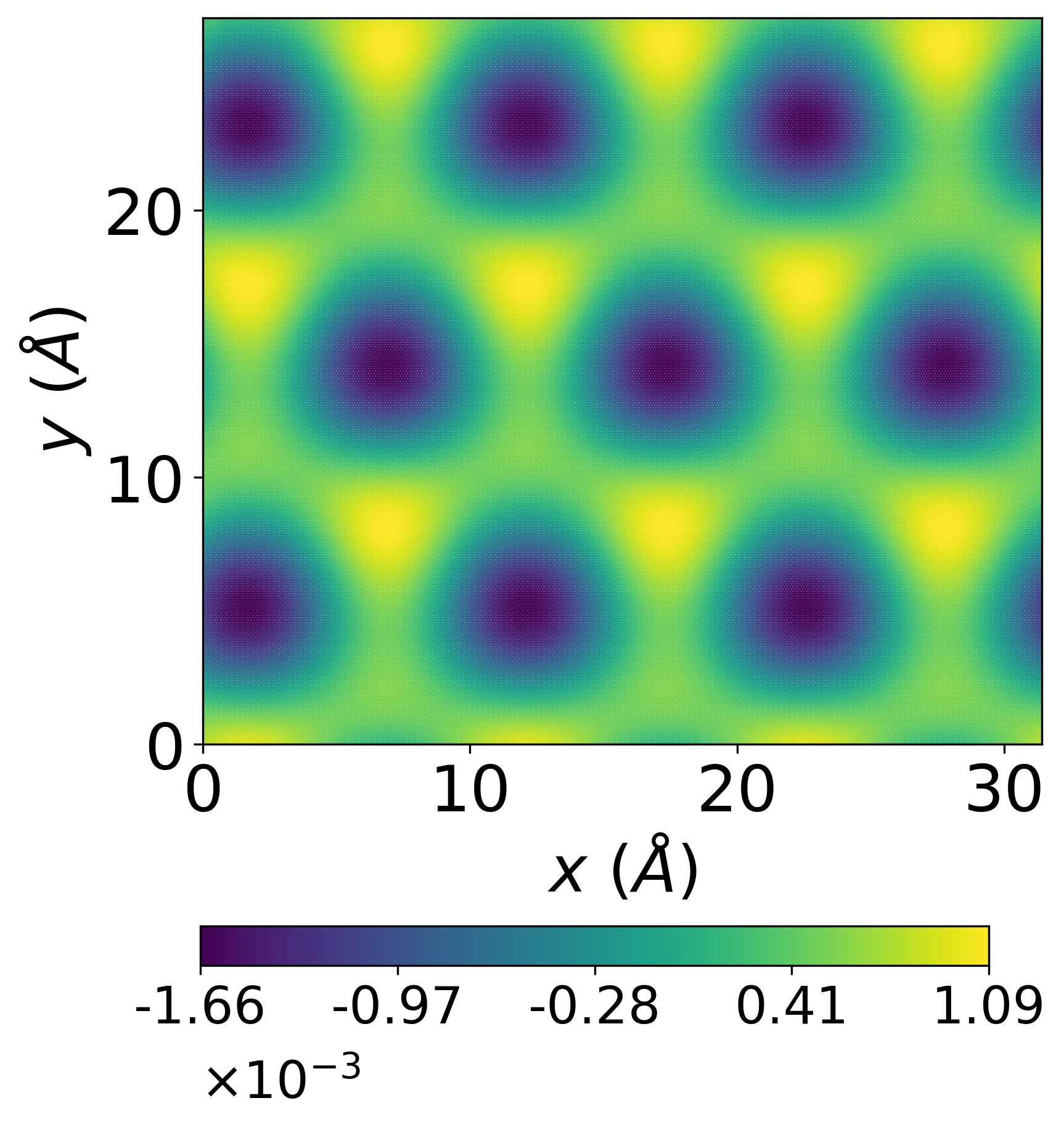}
    \end{subfigure}%
    ~
    \begin{subfigure}[h]{\cdwscaleb}
        \centering
        \caption{}
        \includegraphics[width=\textwidth]{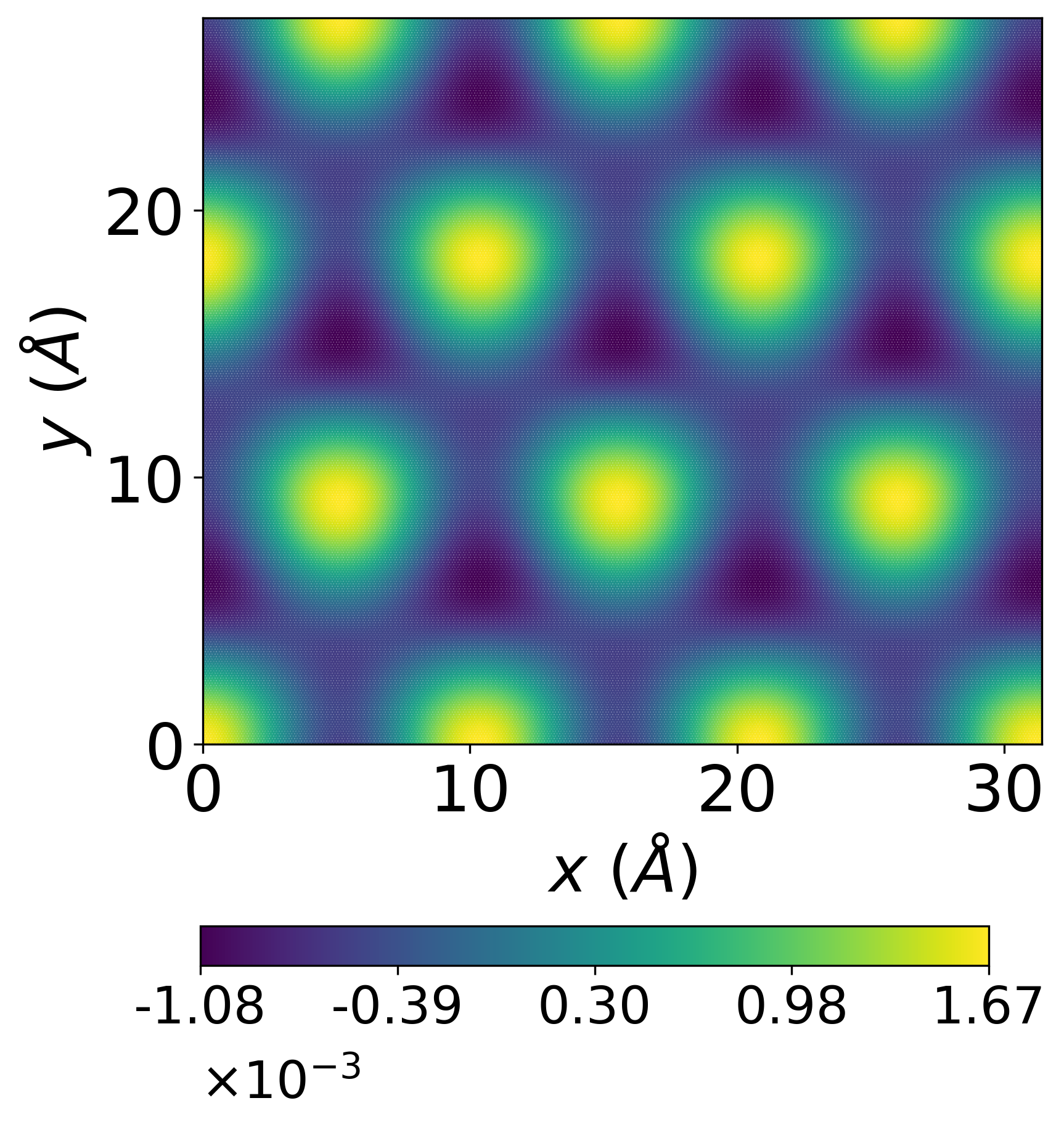}
    \end{subfigure}%
    \caption{Top panels: atomic displacements (orange arrows) of Nb atoms (red circles) in different CDW states. Bottom panels: corresponding smeared Nb density in units of $~\Ang^{-2}$. Black circles indicate Se atoms. For clarity, we show the difference between the smeared Nb density and its mean value.}
    \label{fig:at_displ_den}
\end{figure*}

In high-symmetry monolayer NbSe$_2$, there are three soft phonon modes with wavevectors $\wavevec_1 = \frac{2\pi}{3\sqrt{3}\lattconst}(1,  -\sqrt{3})$, $ \wavevec_2 = \frac{2\pi}{3\sqrt{3}\lattconst}(1, \sqrt{3})$ and $\wavevec_3 = \frac{2\pi}{3\sqrt{3}\lattconst} (-2, 0)$. These soft modes give rise to CDW formation in the monolayer. The corresponding atomic displacements are given by  
\begin{equation}
    \delta\atpos_i=\sum_{j=1}^3\displamp_j\hat{\wavevec}_j\cos(\wavevec_j\cdot\atpos_i+\displphase_j), 
    \label{eq:cdw_displ}    
\end{equation}
where $\displamp_j$ and $\displphase_j$ are the displacement amplitudes and the phases of displacement along the CDW wavevectors, respectively~\cite{jw-cdw1}. The Nb displacements of the hollow-center, filled-center, and hexagonal CDWs are shown in Figure~\ref{fig:at_displ_den}(a),~(b), and~(c) respectively. Each CDW type is characterized by a specific combination of phases $\displphase_j$ which are summarized in Table~\ref{tab:displphi_type} of the Supplementary Materials. 

Similar to the relaxed atomic positions, the smeared Nb atom density of the NbSe$_2$ monolayer in the CDW state can also be approximated as a Fourier series according to
\begin{equation} \fitden(\pos) = \fitden_0 + \sum_{j=1}^{3} \fitden_{j} \cos(\wavevec_j\cdot\pos+ \phase_j ), 
    \label{eq:rho_ft_const}
\end{equation}
where $\fitden_0$ is the mean atomic density, $\fitden_{j}$ is the density oscillation amplitude along $\wavevec_j$, and $\phase_j$ is the phase of the oscillation along $\wavevec_j$. 

For uniform CDWs, the amplitudes of the density oscillation ($\fitden_j$) are proportional to the magnitude of the atomic displacements ($\displamp_j$), while the phases of the density oscillation ($\phase_j$) characterize the displacement patterns. This is because the values of $\phase_j$ determine the positions of the maxima in the smeared atomic density relative to some chosen origin. For the filled-center, hollow-center, and hexagonal CDWs, the density maxima are located at Se, hollow, and Nb sites, respectively. Each CDW is therefore associated with a distinct combination of phases. Conversely, one can use the phases to identify the CDW type. 

The smeared Nb atom density of the hollow-center CDW, computed using $\smearing=0.85\lattconst$, is shown in Figure~\ref{fig:at_displ_den}(d). The observed modulation of the smeared atomic density can be viewed as a standing wave obtained by superposing the three density oscillations along each wavevector. The smeared Nb atom densities of the filled-center and hexagonal CDWs are shown in Figures~\ref{fig:at_displ_den}(e) and~(f), respectively. Similar interference patterns are observed, but the density maxima and minima have different positions compared to that of the hollow-center CDW. 

In twisted bilayers, different types of CDWs can co-exist in the moir\'e unit cell~\cite{cdw_nbse2_cheung}. This is because the different CDW types have very similar energies and small changes in the local strain can give rise to a different energetic ordering. In our calculations, the hollow-center CDW is most stable but the total energies (per $3\times3$ unit cell) of the filled-center and hexagonal CDWs are only about 20-30 meV higher, see Table~\ref{tab:cdw_enr} of Supplementary Materials. To describe such a twisted bilayer system, we allow the mean density, the amplitudes and phases of the smeared atomic density to vary over space according to
\begin{equation} \fitden(\pos)=\fitden_0(\pos)+\sum_{j=1}^{3}\fitden_{j}(\pos)\cos(\wavevec_j\cdot\pos+\phase_j(\pos)). 
    \label{eq:rho_ft}
\end{equation}

To determine the $\fitden_j(\pos)$ and $\phase_j(\pos)$ from a given smeared atomic density corresponding to a relaxed atomic structure, we proceed as follows: first, the three phase fields $\phase_j(\pos)$ are obtained by repeatedly projecting gradients of the smeared metal atom density onto unit vectors $\hat{\mathbf{n}}_j$ that are orthogonal to the corresponding CDW wavevector $\mathbf{q}_j$ (see Section~\ref{sec:lpa_derivation} of the Supplementary Materials for the derivation). For example, $\phase_1(\pos)$ is extracted from
\begin{equation}
        \tan(\wavevec_1 \cdot \pos+\phase_1(\pos))=-\frac{1}{|\wavevec_1|^2}\frac{\wavevec_1\cdot\mathbf{\nabla}(\hat{\normvec}_2\cdot\mathbf{\nabla}(\hat{\normvec}_3\cdot\mathbf{\nabla}\den(\pos)))}{\hat{\normvec}_2\cdot\mathbf{\nabla}(\hat{\normvec}_3\cdot\mathbf{\nabla}\den(\pos))}.
    \label{eq:phi_tan}
\end{equation}
From the values of these phases, the local CDW type can be identified. The type of CDW is classified by checking whether the local atomic density maximum is located near an interstitial site, a Nb atom, or a Se atom below a certain distance along all three CDW wavevectors, see Section~\ref{sec:lpa_derivation} of the Supplementary Materials for additional details~\footnote{
We note that there is an image-processing technique for obtaining phase profiles~\cite{visualize_triangular_cdw}. It is a lock-in technique that involves filtering the smeared atomic density image. However, Equation~\ref{eq:phi_tan} enables local phase extraction from the positions of a small number of Nb atoms around a point. It demonstrates that the phases can be obtained from the projected curvatures, which are (semi-)local quantities. It eliminates the need to compute the global smeared atomic density image when obtaining the phase profile. See Section~\ref{sec:derivsden} of Supplementary Materials for how to obtain the phases from the Nb positions.}. In the twisted bilayer, we found that the CDWs are highly deformed compared to the uniform $3\times3$ CDWs shown in Figure~\ref{fig:at_displ_den}. This is also observed in ~\cite{cdw_nbse2_cheung}. Our method thus identifies the $3\times3$ uniform CDW that best resembles the deformed CDW based on the position of the local atomic density maximum. The identification results could be different from identification methods based on the local bond lengths or the atomic displacements~\cite{cdw_nbse2_cheung}.  

Next, the mean density $\fitden_0(\pos)$ and the density oscillation amplitudes $\fitden_j(\pos)$ are determined numerically by minimizing the Lagrangian
\begin{equation}
    \Lagr=\int(\fitden(\pos)-\den(\pos))^2 d^2\pos. 
    \label{eq:amp_L}
\end{equation}
Note that $\fitden(\pos)$ denotes the approximated atomic density defined in Equation~\ref{eq:rho_ft} and $\den(\pos)$ denotes the exact smeared atomic density defined in Equation~\ref{eq:rho_gauss}. By minimizing this Lagrangian, the parameters $\{\fitden_j\}$ that give the best fit to the the actual smeared atomic density are determined. Specifically, the four Euler-Lagrange equations representing the variation of $\Lagr$ with respect to $\fitden_0(\pos)$ and $\fitden_j(\pos)$ are derived and solved using the steepest descent method. This allows us to deduce the ``CDW field'' in the given layer. More details of the implementation is included in Section~\ref{sec:el_amp} in the Supplementary Materials.

\section{Results}
Figure~\ref{fig:cdw_tb}(a) shows the smeared Nb density $\den(\pos)$ for $\smearing=0.85\lattconst$ for the twisted NbSe$_2$/MoSe$_2$ bilayer. The smeared Nb density exhibits a modulation characteristic of the formation of a CDW. In contrast to the smeared Nb density of the monolayer shown in Fig.~\ref{fig:at_displ_den}, the smeared Nb density of the twisted bilayer does not exhibit a regular modulation pattern throughout the moir\'e unit cell, indicating the co-existence of different CDW types. \footnote{We note that the smeared atomic density does not exhibit a three-fold rotation symmetric structure as was observed in our recent work in twisted bilayer NbSe$_2$~\cite{cdw_nbse2_cheung}. We were able to find a structure that exhibits a symmetric smeared atomic density but has a slightly higher total energy. We analyze this metastable structure in Section~\ref{sec:metastable_struct} of the Supplementary Materials, but note that all qualitative findings obtained for the non-symmetric low-energy structure are also found in the high-symmetry structure.}

To identify which CDW types co-exist in the NbSe$_2$ layer of the twisted bilayer, we calculate the phase fields $\phase_j(\pos)$ and compare their values to those of the uniform CDW phases of the monolayer. Fig.~\ref{fig:cdw_tb}(c) shows that filled-center, hollow-center and hexagonal CDWs are all observed. In particular, a large area of hollow-center CDW is found around the MM center of the unit cell. This region is surrounded by smaller patches of filled-center CDWs and hexagonal CDWs. We have previously observed a similar co-existence of different CDW types in a twisted homo-bilayer of NbSe$_2$~\cite{cdw_nbse2_cheung} and discussed in detail how local strain induced by moir\'e relaxations controls the relative stability of these phases. 

We now turn our attention to the MoSe$_2$ layer of the NbSe$_2$/MoSe$_2$ twisted bilayer. Figure~\ref{fig:cdw_tb}(b) shows the smeared Mo density for $\smearing=0.85\lattconst$. Even though monolayer MoSe$_2$ does not exhibit a CDW, the smeared Mo density exhibits a modulation characteristic of CDW formation. However, the amplitude of the CDW modulations is approximately an order of magnitude smaller than in the NbSe$_2$ layer. Comparing the smeared Mo density to the smeared Nb density reveals that maxima in the Mo density are aligned with top of minima of the Nb density and vice versa. For example, the Nb density exhibits a maximum at the MM center, while the Mo density exhibits a minimum, see also Figure~\ref{fig:compare_extr_den} of the Supplementary Materials.

Again, different types of CDWs can be identified in the MoSe$_2$ layer, as shown in Figure~\ref{fig:cdw_tb}(d). Similarly to the NbSe$_2$ layer, a large region of hollow-center CDW is observed in the vicinity of the MM center which is surrounded by smaller regions of filled-center and hexagonal CDWs.

\newcommand{\alphaFactor}{0.95}    
\newcommand{\xshift}{-14pt}
\newcommand{\captionhspacea}{85pt}
\newcommand{\captionhspaceb}{95pt}

\definecolor{pythoncyan}{HTML}{00FFFF}
\definecolor{pythonorange}{HTML}{FFA500}
\definecolor{pythonpurple}{HTML}{800080}

\begin{figure*}[htb!]
    \begin{subfigure}[t]{0.49\textwidth}
        \centering
        \caption{\hspace{\captionhspacea}NbSe$_2$ layer}
        \includegraphics[width=\textwidth]{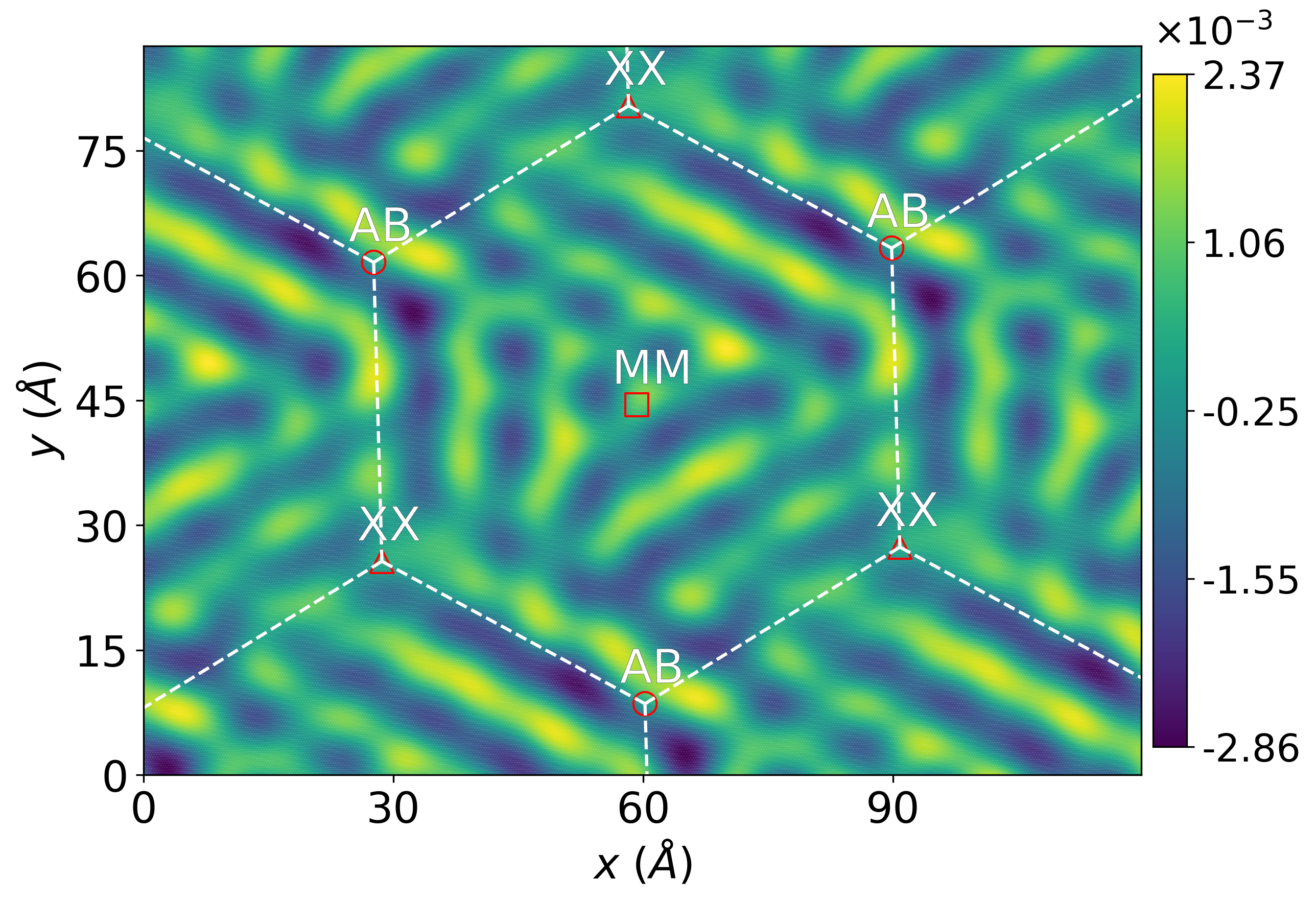}
    \end{subfigure}%
    ~  
    \begin{subfigure}[t]{0.49\textwidth}
        \centering
        \caption{\hspace{\captionhspacea}MoSe$_2$ layer}
        \includegraphics[width=\textwidth]{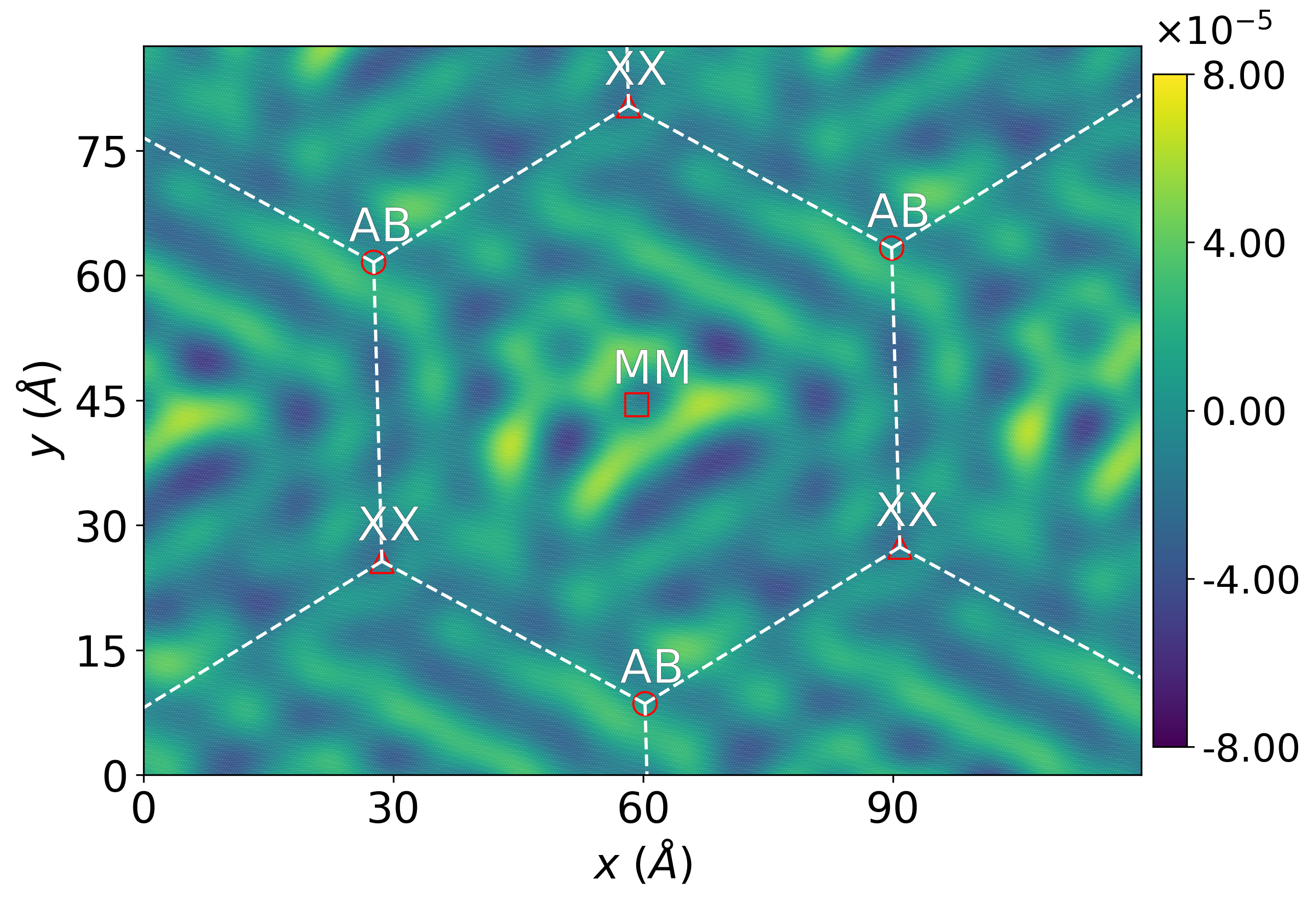}
    \end{subfigure}
    ~
    \begin{subfigure}[t]{0.49\textwidth}
         \centering
         \caption{\hspace{\captionhspaceb}NbSe$_2$ layer}
         \hspace{\xshift}
         \includegraphics[width=\alphaFactor\textwidth]{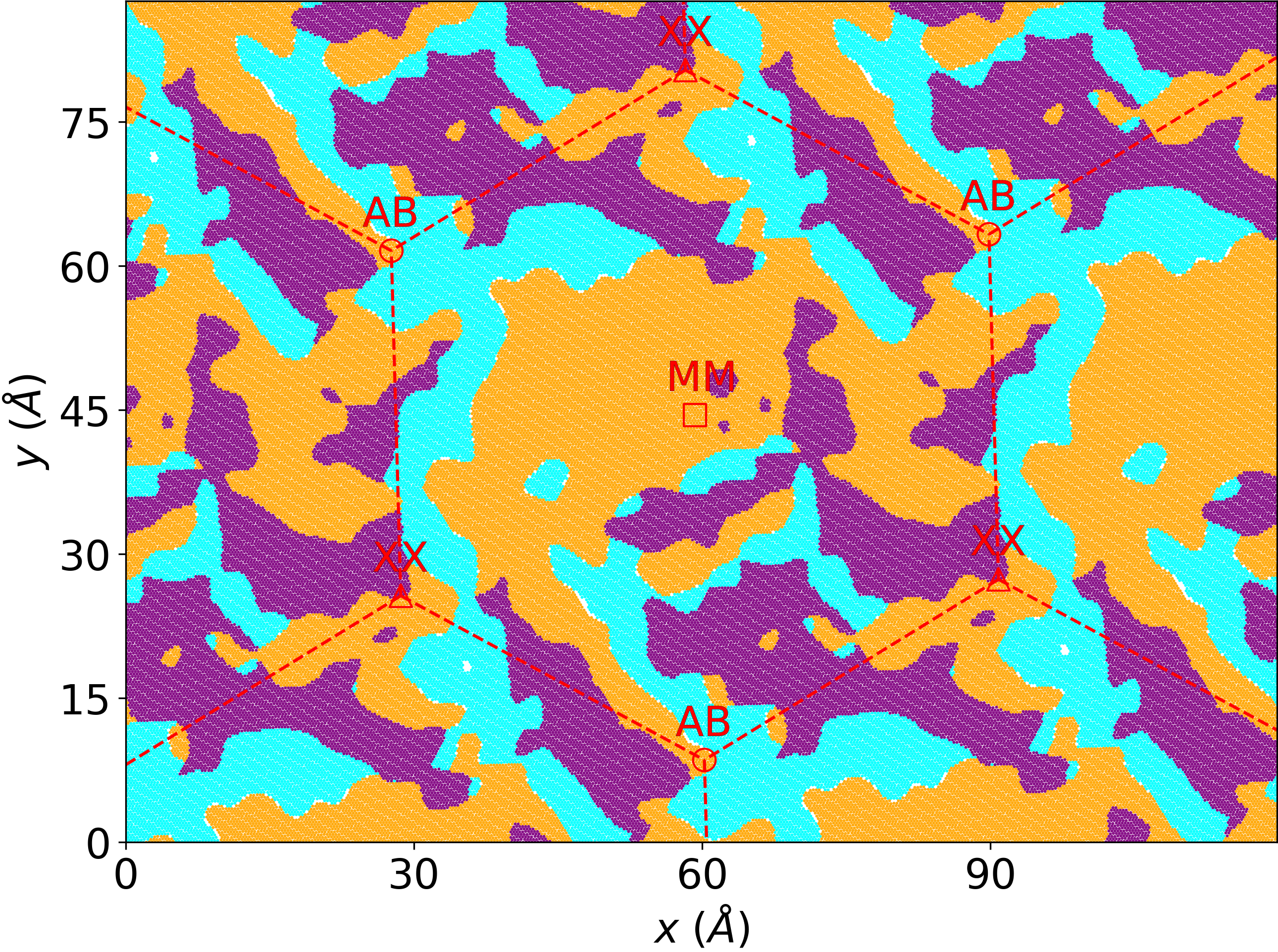}
     \end{subfigure}%
    ~
    \begin{subfigure}[t]{0.49\textwidth}
         \centering
         \caption{\hspace{\captionhspaceb}MoSe$_2$ layer}
         \hspace{\xshift}
         \includegraphics[width=\alphaFactor\textwidth]{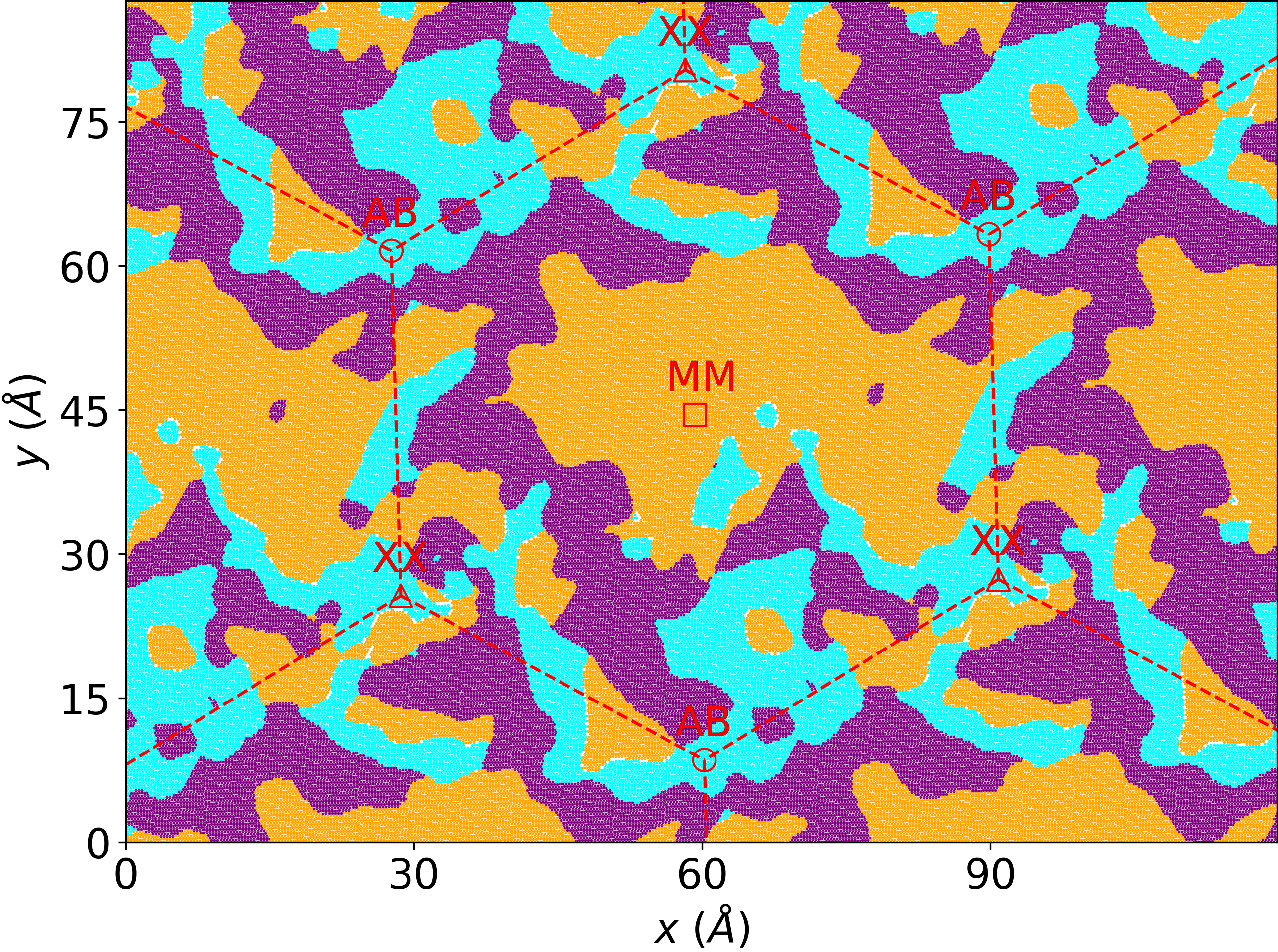}
    \end{subfigure}%

    \vspace{10pt}
    Legend:~\colorbox{pythonorange}{Hollow-center CDW}    \colorbox{pythoncyan}{Filled-center CDW} \colorbox{pythonpurple}{\textcolor{white}{Hexagonal CDW}}
    
    \caption{Smeared atomic densities (in units of $~\Ang^{-2}$) of Nb atoms (a) and Mo atoms (b) in a 3.15$^\circ$ twisted MoSe$_2$/NbSe$_2$ bilayer. For clarity, the difference between the smeared atomic density and its mean value is shown. Centers of high-symmetry stacking regions are marked in red, and the Wigner-Seitz cell of the moir\'e superlattice is indicated in white. Maps of different CDW types in the NbSe$_2$ layer (c) and the MoSe$_2$ layer (d). Centers of high-symmetry stacking regions are marked in black, and the Wigner-Seitz cell of the moir\'e superlattice is indicated in black. }
    \label{fig:cdw_tb}
\end{figure*}

\begin{figure*}[htbp!]
    \begin{subfigure}[t]{0.49\textwidth}
        \centering
        \caption{\hspace{\captionhspacea}NbSe$_2$ layer}
        \includegraphics[width=\textwidth]{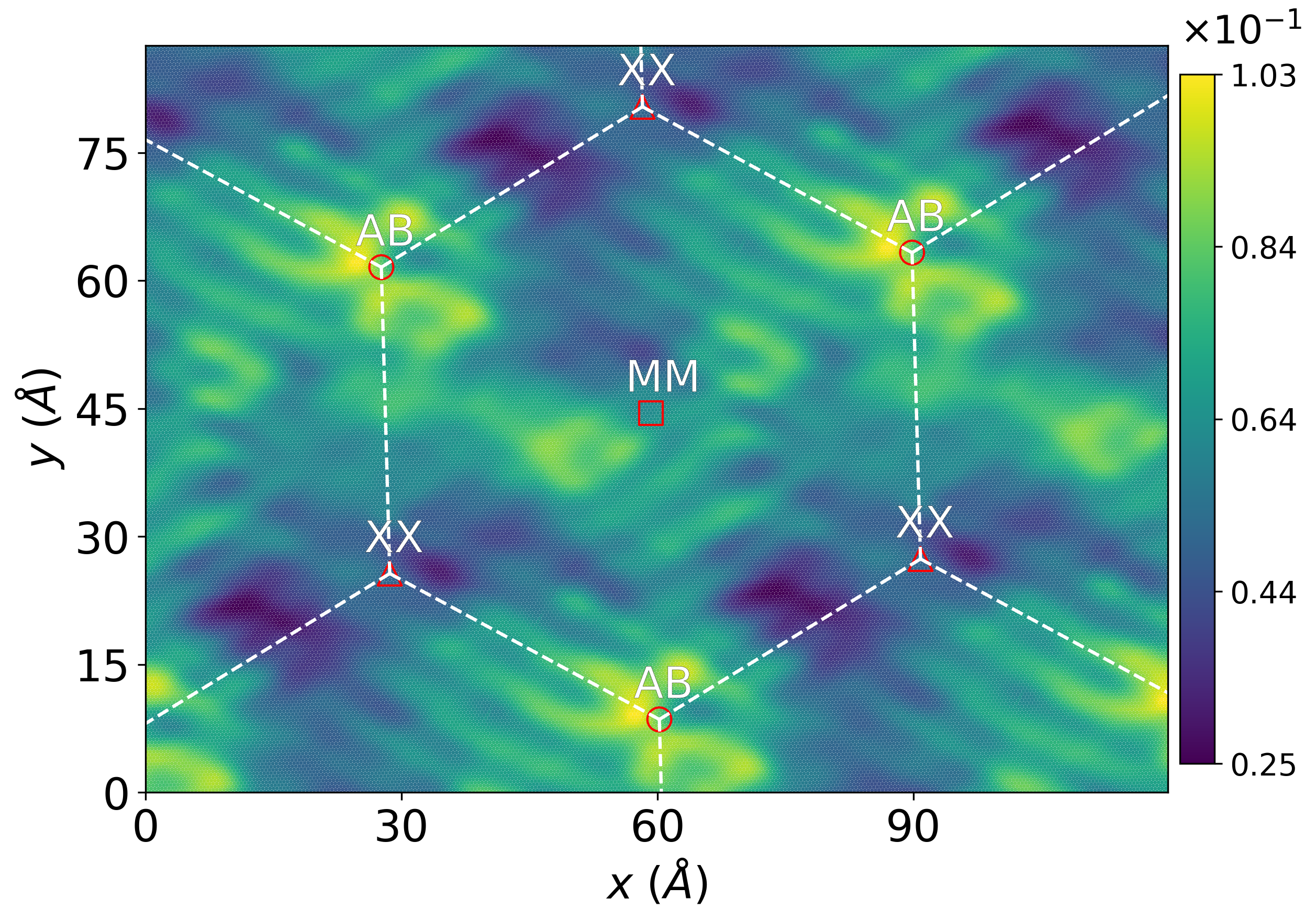}
    \end{subfigure}
    ~  
    \begin{subfigure}[t]{0.49\textwidth}
        \centering
        \caption{\hspace{\captionhspacea}MoSe$_2$ layer}
        \includegraphics[width=\textwidth]{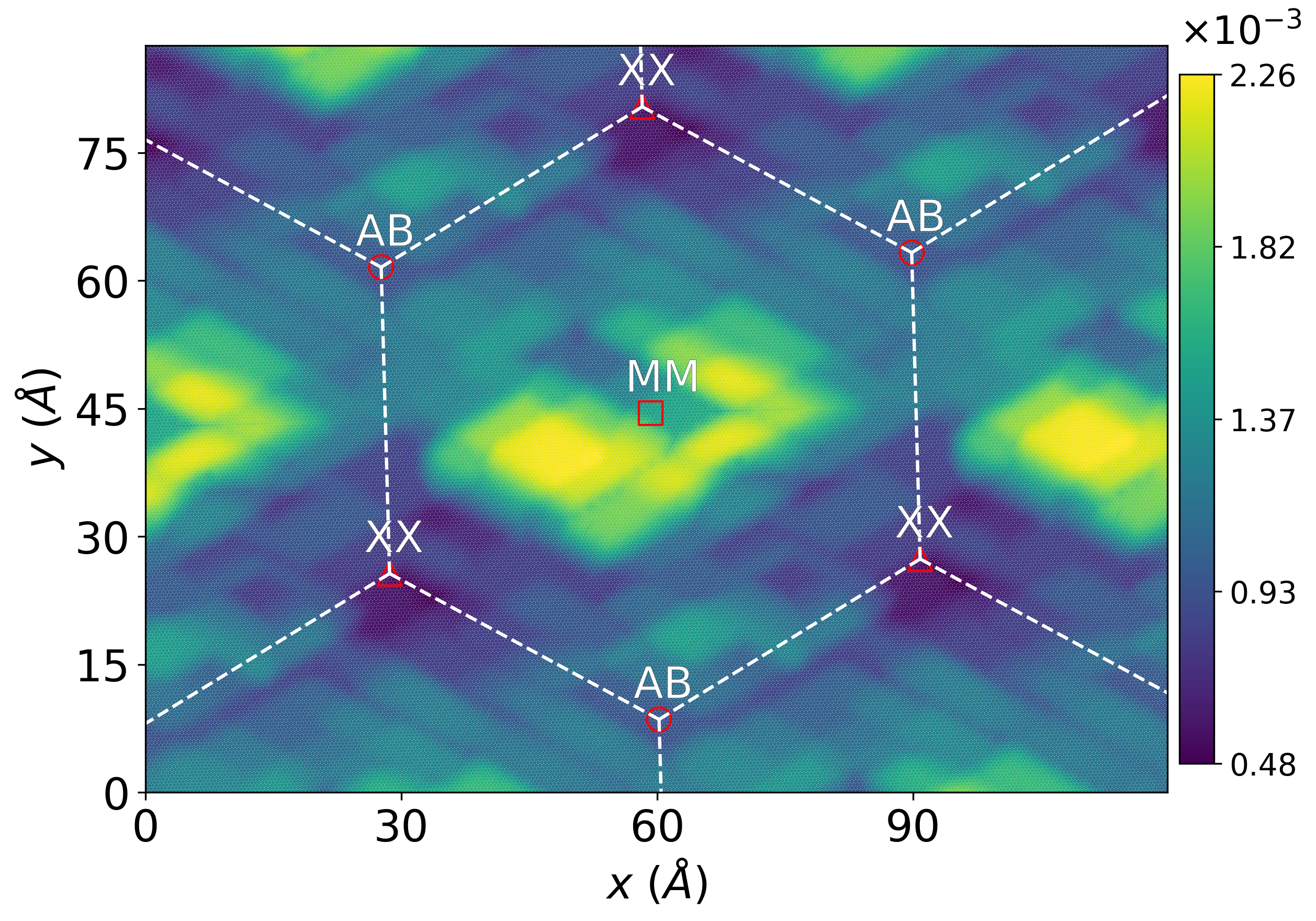}
    \end{subfigure}
    \caption{The average CDW displacement amplitude (in $\Ang$) of (a) Nb atoms and (b) Mo atoms in a twisted MoSe$_2$/NbSe$_2$ bilayer. Centers of high-symmetry stacking regions are marked in red, and the Wigner-Seitz cell of the moir\'e superlattice is indicated in white.}
    \label{fig:amp_tb}
\end{figure*}

Next, we analyze the CDW displacement amplitudes in more detail, see Fig.~\ref{fig:amp_tb}. In the NbSe$_2$ layer, the CDW displacement amplitudes at the MM, AB, and XX centers are $6.5\times10^{-2}~\Ang$, $7.7\times10^{-2}~\Ang$, and $4.8\times10^{-2}~\Ang$, respectively. In contrast, the CDW displacement amplitudes are about an order of magnitude smaller in the MoSe$_2$ layer: $1.8\times10^{-3}~\Ang$ in the MM stacking region; $1.1\times10^{-3}~\Ang$ in the AB region; and $7.4\times10^{-4}~\Ang$ in the XX region. Interestingly, we find that the strongest CDW is found in the AB region in the NbSe$_2$ layer, but in the MM region for the MoSe$_2$ layer. 

\begin{figure}[htbp!]
    \centering
    \includegraphics[width=\linewidth]{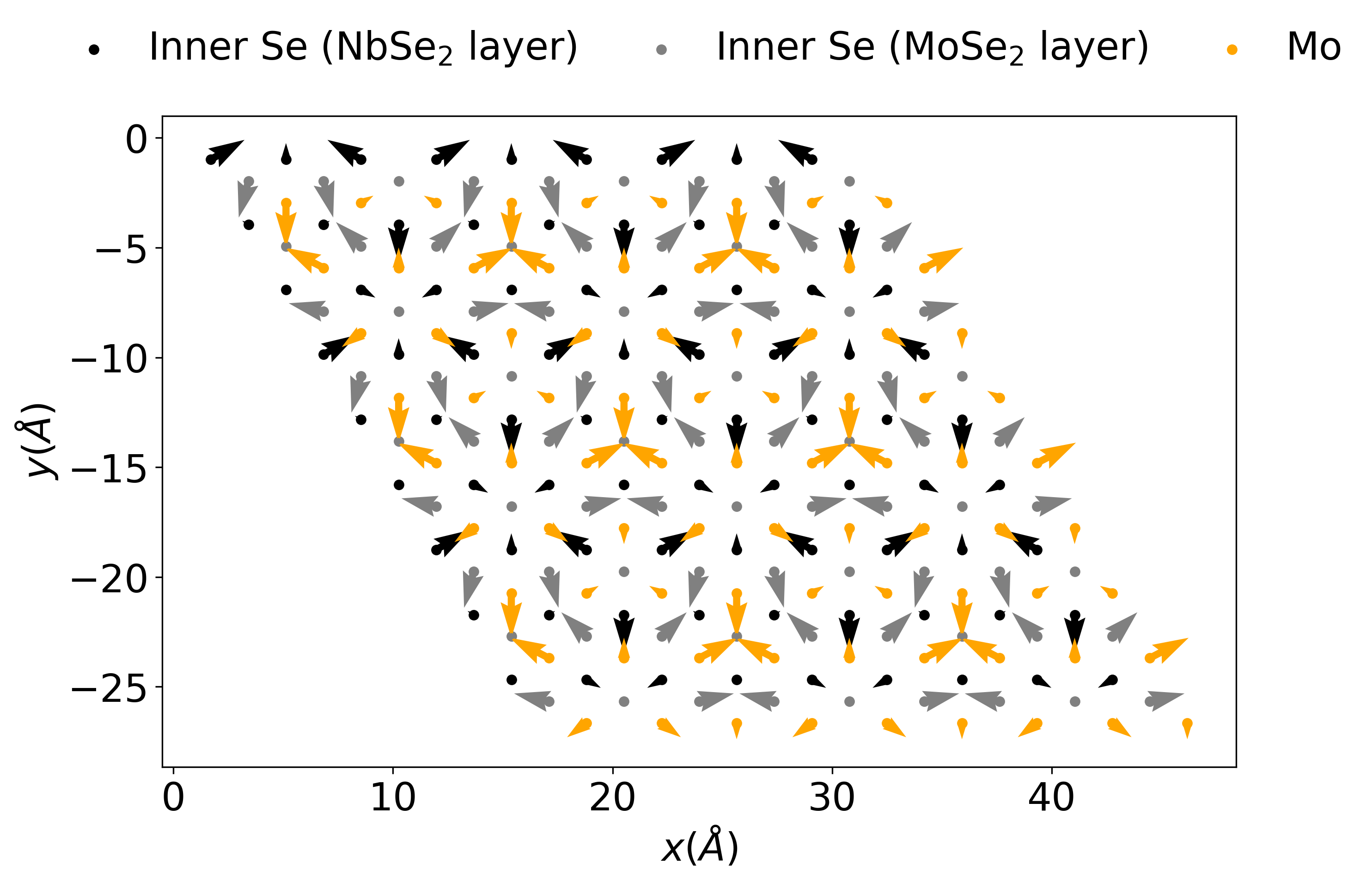}
    \caption{Atomic displacements of the inner Se atoms in the NbSe$_2$ layer (black), MoSe$_2$ layer (grey), and Mo atoms (orange) in an untwisted MM-stacked bilayer of NbSe$_2$/MoSe$_2$. The displacements of the inner Se atoms in the NbSe$_2$ layer are multiplied by a factor of 50, while the displacements of the inner Se and Mo atoms in the MoSe$_2$ are multiplied by a factor of 750.}
    \label{fig:displ_mm}
\end{figure}

To understand these observations and identify the microscopic mechanism driving CDW formation in the MoSe$_2$ layer, we next study untwisted NbSe$_2$/MoSe$_2$ bilayers with AB, XX, and MM stacking. Specifically, we relax the three untwisted bilayers using a $3\times3$ supercell and then inspect the atomic displacement vectors in both layers, see Fig.~\ref{fig:displ_mm} for the case of an MM stacked untwisted bilayer. We focus on the displacement vectors of the inner Se atoms of each layer and of the Mo atoms. It can be observed that the inner Se atoms of the MoSe$_2$ layer (grey circles) tend to move in the opposite direction to the Se atoms of the NbSe$_2$ layer (black circles), while the Mo atoms (orange circles) move in the same direction as the Se atoms they are bonded to. The same behavior is found for the XX and AB bilayers, see Figure~\ref{fig:cdw_displ_highsym_all} of the Supplementary Materials. This suggests that the formation of CDWs in the MoSe$_2$ layer is mediated by the steric repulsion between the inner Se atoms of the two layers: the CDW formation in the NbSe$_2$ layer induces a displacement of the Se atoms which push the inner Se atoms of the MoSe$_2$ layer ``out of the way"; this, in turn, drives the displacements of the Mo atoms which follow the Se atoms to reduce the distortion of the intralayer Mo--Se bonds. As a consequence of this correlated movement, Mo atoms avoid positions directly across the van der Waals gap from Nb atoms explaining why maxima of the Mo smeared atomic density coincide with minima of the Nb smeared atomic density, see Figure~\ref{fig:compare_extr_den} of the Supplementary Materials. 

\begin{figure}[htbp!]
   \centering
   \includegraphics[width=\linewidth]{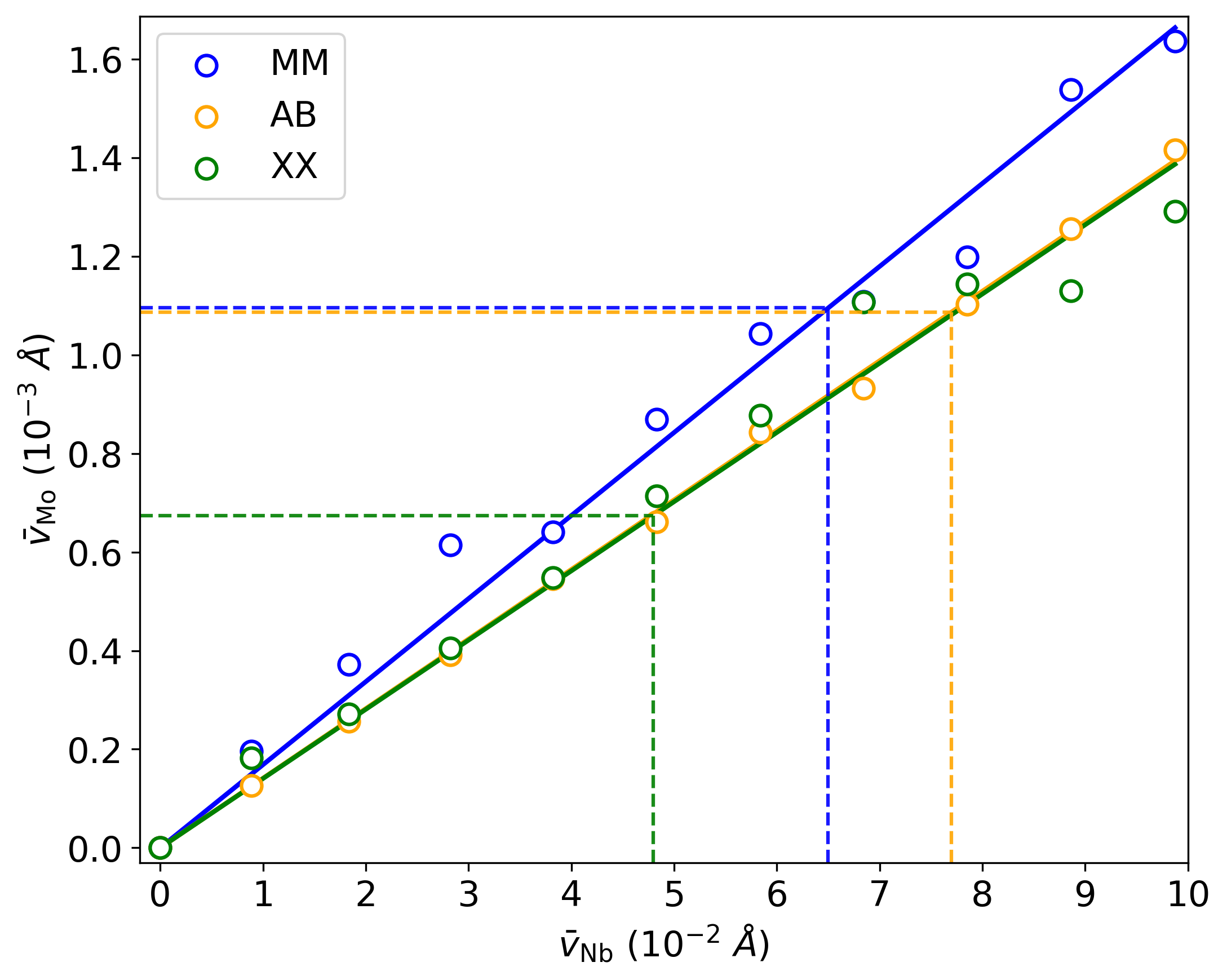}
   \caption{CDW displacement amplitude in the MoSe$_2$ layer as function of the CDW displacement amplitude in the NbSe$_2$ layer for different untwisted bilayers. In these calculations, a hollow-centered CDW is imposed on the Nb atoms and all other atoms are allowed to relax. The solid lines of best linear fits for each stacking. Dashed vertical lines indicate the NbSe$_2$ CDW amplitude extracted in the corresponding stacking regions of the twisted bilayer and dashed horizontal lines indicate the resulting predictions for the MoSe$_2$ CDW amplitude.}
   \label{fig:amp_amp}
\end{figure}

Based on the above discussion, we expect a reduction (increase) of the CDW amplitude in the NbSe$_2$ layer to result in a corresponding reduction (increase) in the CDW amplitude of the MoSe$_2$. To establish whether this is indeed the case, we perform calculations of untwisted bilayers in which we fix the hollow-center CDW amplitude in the NbSe$_2$ layer. Figure~\ref{fig:amp_amp} shows that for all bilayers the amplitude of the induced CDW in the MoSe$_2$ layer increases approximately linearly with the CDW amplitude in the NbSe$_2$ layer. Interestingly, our calculations reveal that the fastest increase in the MoSe$_2$ CDW amplitude is found for the MM-stacked bilayer, while the slopes are similar for AB- and XX-stacked bilayers. In particular, we have used the linear fits obtained from the untwisted bilayer calculations (solid lines in Fig.~\ref{fig:amp_amp}) and the strengths of the NbSe$_2$ CDW amplitude at high-symmetry stackings of the twisted bilayer (dashed vertical lines) to predict the corresponding CDW amplitude in the MoSe$_2$ layer (dashed horizontal lines). 

This procedure yields accurate quantitative agreement for the MoSe$_2$ CDW amplitude in the XX and AB stacking regions: for the XX stacking, the predicted amplitude is $6.9\times10^{-4}~\Ang$ compared to a value of $7.4\times10^{-4}~\Ang$ extracted from the twisted bilayer calculation. For the AB stacking, the predicted amplitude is $1.1\times10^{-3}~\Ang$ which exactly matches the value obtained from the explicit calculation. However, for the MM stacking, the extracted value from the twisted bilayer calculation ($1.8\times10^{-3}~\Ang$) is significantly larger than the one expected from the untwisted bilayer ($1.1\times10^{-3}~\Ang$).

To understand this enhancement in the MM region of the twisted bilayer, we inspected the distribution of charge in the moir\'e unit cell in both layers (see Supplementary Figure~\ref{fig:q_inplane}) and also calculated the magnitude of interlayer charge transfer in the untwisted bilayers (see Supplementary Figure~\ref{fig:amp_ct}). We find that in both twisted and untwisted bilayers electrons are transferred from the MoSe$_2$ layer to the NbSe$_2$ layer. Furthermore, in the twisted bilayer, the transferred electrons localize in the MM region of the NbSe$_2$ layer so that this region acquires negative charge than the corresponding untwisted bilayer. The resulting electric field couples to the negative charge of the inner Se atoms of the MoSe$_2$ layer which in turn results in an increase of the CDW amplitude in the MoSe$_2$ layer.

Finally, we note that doping can also stabilize CDWs in semiconducting TMD monolayers~\cite{mos2_cdw}. However, it is not the dominant mechanism for inducing CDWs in bilayer NbSe$_2$/MoSe$_2$: although the MoSe$_2$ layer is indeed hole-doped due to interlayer charge transfer, we found that the hole concentration is too small to stabilize a CDW in the MoSe$_2$ layer (see Section~\ref{sec:ct_utb} of Supplementary Materials). 

\section{Conclusions}
In this work, we demonstrated that CDWs are induced in the semiconducting MoSe$_2$ layer through a proximity effect in a twisted NbSe$_2$/MoSe$_2$ bilayer. We reveal steric repulsion between Se atoms across the van der Waals gap as the microscopic mechanism driving CDW formation in the MoSe$_2$ layer. As this coupling is highly sensitive to the stacking, the amplitude of the induced CDW varies on the moir\'e length scale. Furthermore, the localization of electrons in the NbSe$_2$ layer enhances the amplitude of the induced CDWs in the MoSe$_2$ layer compared that achievable in untwisted bilayers. We expect this enhancement to be even more pronounced at smaller twist angles. We note that this effect could potentially be used to control superconductivity in the MoSe$_2$ layer~\cite{mos2_cdw}. Our work shows that it is possible to engineer symmetry-broken states in 2D materials by constructing twisted heterobilayers, and potentially offer a route to induce superconducting states in these systems. 

\section{Acknowledgements}
C. T. S. C. acknowledges funding from the Croucher Foundation and an Imperial College President's Scholarship. This work used the ARCHER2 UK National Supercomputing Service
(https://www.archer2.ac.uk) and the Imperial College Research Computing Service facility CX1 (DOI: 10.14469/hpc/2232). We acknowledge the Thomas Young Centre under Grant No. TYC-101.

\clearpage

\makeatletter
\def\@seccntformat#1{%
  \csname the#1\endcsname\quad
}
\makeatother

\onecolumngrid
\section{Supplementary Materials}
\setcounter{secnumdepth}{3}

\renewcommand{\figurename}{SUPPLEMENTARY FIG.}
\renewcommand{\thefigure}{S\arabic{figure}}
\renewcommand{\theequation}{S\arabic{equation}}
\renewcommand{\thetable}{S\arabic{table}}
\renewcommand{\thesection}{S\arabic{section}}
\renewcommand{\thesubsection}{S\arabic{section}.\arabic{subsection}}
\renewcommand{\thesubsubsection}{S\arabic{section}.\arabic{subsection}.\arabic{subsubsection}}

\setcounter{section}{0}
\setcounter{subsection}{0}
\setcounter{subsubsection}{0}
\setcounter{table}{0}
\setcounter{figure}{0}
\setcounter{equation}{0}

\section{Methods}
\label{sec:methods}

\subsection{Construction of the moir\'e unit cell}
\label{subsec:moire}

To construct the initial positions for the atomic relaxations, we stack and twist two flat anti-parallel NbSe$_2$ and MoSe$_2$ monolayers that do not feature a charge density wave. The resulting moir\'e unit cell is spanned by the lattice vectors $\vc{t}_1$ and $\vc{t}_2$ given by 
\begin{align}
    \notag
    \vc{t}_1 &= n\vc{a}_1 + m\vc{a}_2, \\
    \vc{t}_2 &= -m\vc{a}_1 + (n+m)\vc{a}_2,
    \label{eq:moiret}
\end{align}
where $\vc{a}_1=a_0/2(\sqrt{3},1,0)$ and $\vc{a}_2=a_0/2(\sqrt{3},-1,0)$, and $a_0$ is the lattice constant. 

The integers $n$, $m$ are the number of unit cells along each moir\'e lattice vector. The twist angle is given by $\cos{\theta}=(n^2+4nm+m^2)/[2(n^2+nm+m^2)]$~\cite{moire_def_dft}. Thus, choosing the pair of integers $(n,m)$ as $(10,11)$ gives the twist angle of $3.15\degree$. The number of atoms in the moir\'e unit cell is $N=6(n^2+nm+m^2)$, which is 1,986 at the chosen twist angle. 

\subsection{Computational details}
\label{subsec:comp_details}

The large-scale ab initio density functional theory code SIESTA~\cite{Soler_Artacho_Gale_Garcia_Junquera_Ordejon_Sanchez-Portal_2002} is used for relaxation and electronic structure calculations. To capture van der Waals interactions, the exchange-correlation functional optB88-vdw is used~\cite{KBM}. We use Troullier-Martins pseudopotentials~\cite{TMPP}, and a double-$\zeta$ polarized basis where the ``split norm" is 0.583 and the energy shift is 0.0109 Ry. We use $\Gamma$-point sampling of the first Brillouin zone. In the self-consistent cycle, the Hamiltonian and the density matrix are converged to 1~$\mu$eV and $1\times 10^{-7}$ per atom, respectively. The force tolerance is set to $3\times 10^{-2}~\text{eV}/\Ang$. The electronic temperature is set to 15.7 K, which is equal to 0.1 mRy. The mesh grid cutoff is 500 Ry. 

In the relaxation of the unit cells of the untwisted bilayers of NbSe$_2$/MoSe$_2$, a Monkhorst-Pack grid of $32\times23\times1$ is used for the k-point sampling. For the $3\times3$ cell relaxations, a Monkorst-Pack grid of $10\times10\times1$ is used. All other parameters are the same as the twisted bilayer relaxation. For the simulation cell, the unit vectors $\vc{a}'_1=3a_0(1,0,0)$ and $\vc{a}'_2=3a_0/2(1,-\sqrt{3},0)$ are used.

\subsection{Smeared atomic density and its Fourier representation}
\label{subsec:dengaussft}
We now show that the smeared atomic density defined in Equation~\ref{eq:rho_gauss} becomes a Fourier series generated by the three charge density wavevectors when the smearing is comparable to the lattice constant. 

Suppose $\atpos_i$ represents the initial position of the $i$-th atom in the high-symmetry lattice before any displacement, and is displaced according to the displacement field $\atpos_i\rightarrow\atpos_i+\delta\atpos_i$, where $|\delta\atpos_i|<<|\atpos_i-\atpos_j|$ for any $i\neq j$, and $|\delta\atpos_i|<<\smearing$ , then the smeared atomic density transforms as,
\begin{align}
    \nonumber
    \den(\pos) &\rightarrow \den'(\pos) \\ 
    \nonumber
    &= \sum_{i=1}^N\exp\left(-\frac{1}{2\smearing^2}(\pos-(\atpos_i+\delta\atpos_i))^2\right)\\
    \nonumber
    &=\sum_{i=1}^N\exp\left(-\frac{1}{2\smearing^2}(\pos-\atpos_i)^2\right)\exp\left(-\frac{1}{\smearing^2}(\pos-\atpos_i)\cdot\delta\atpos_i\right)\exp\left(-\frac{1}{2\smearing^2}\delta\atpos_i^2\right)\\
    \nonumber
    &\approx \sum_{i=1}^N\exp\left(-\frac{1}{2\smearing^2}(\pos-\atpos_i)^2\right)\exp\left(-\frac{1}{\smearing^2}(\pos-\atpos_i)\cdot\delta\atpos_i\right)\\
    \nonumber
    &\approx \sum_{i=1}^N\exp\left(-\frac{1}{2\smearing^2}(\pos-\atpos_i)^2\right)\left[1+\left(-\frac{1}{\smearing^2}(\pos-\atpos_i)\cdot\delta\atpos_i\right)\right] \\
    &= \den(\pos) - \left(\frac{1}{\smearing^2}\sum_{i=1}^N\exp\left(-\frac{1}{2\smearing^2}(\pos-\atpos_i)^2\right)(\pos-\atpos_i)\cdot\delta\atpos_i\right)
    \label{eq:den_change}
\end{align}
We note that since $\atpos_i$ represents a high-symmetry lattice, when a smearing comparable to the lattice constant is used, the associated smeared atomic density is simply the average atomic density, $\den_0$, which is a constant over space. Also, the normalization factor $1/(2\pi\smearing^2)$ is omitted in the above expressions.

Under a charge density wave, the atomic displacements are given by 
\begin{equation}
    \delta\atpos_i=\sum_{j=1}^3\displamp^{(j)}\hat{\wavevec}_j\cos(\wavevec_j\cdot\atpos_i+\displphase_j),
\end{equation}
then Equation~\ref{eq:den_change} becomes
\begin{align}
    \nonumber
    \den'(\pos) 
    &=
    \den_0 - \frac{1}{\smearing^2}\sum_{i=1}^N\exp\left(-\frac{1}{2\smearing^2}(\pos-\atpos_i)^2\right)(\pos-\atpos_i)\cdot \sum_{j=1}^3 \hat{\wavevec}_j \displamp^{(j)}\cos(\wavevec_j\cdot\atpos_i+\displphase_j)\\    
    &=
    \den_0 - \frac{1}{\smearing^2}\sum_{i=1}^N\exp\left(-\frac{1}{2\smearing^2}(\atpos_i')^2\right)\atpos_i' \cdot \sum_{j=1}^3  \hat{\wavevec}_j \displamp^{(j)}\cos(\wavevec_j\cdot(\pos-\atpos_i')+\displphase_j),
\end{align}
by making the change of variable $\atpos_i'=\pos-\atpos_i$. Now 
\begin{align}
    \nonumber
    \den'(\pos) &= 
    \den_0 - \frac{1}{\smearing^2} \sum_{j=1}^3  \displamp^{(j)} \sum_{i=1}^N\exp\left(-\frac{1}{2\smearing^2}(\atpos_i')^2\right) (\atpos_i' \cdot \hat{\wavevec}_j) [\cos(\wavevec_j\cdot\atpos_i')\cos(\wavevec_j\cdot\pos+\displphase_j) \\
    \nonumber
    &+\sin(\wavevec_j\cdot\atpos_i')\sin(\wavevec_j\cdot\pos+\displphase_j)].
    \label{eq:den_coscossinsin}
\end{align}
The sum involving the term $(\atpos_i'\cdot\hat{\wavevec}_j)\cos(\wavevec_j\cdot\atpos_i')$ is odd in $\atpos_i'$ and vanishes if the sum is taken over the entire infinitely-sized crystal. So Equation~\ref{eq:den_coscossinsin} becomes
\begin{align}
    \den'(\pos) 
    &=
    \den_0 -  \sum_{j=1}^3 \left[\frac{\displamp^{(j)}}{\smearing^2} \sum_{i=1}^N\exp\left(-\frac{1}{2\smearing^2}(\atpos_i')^2\right) (\atpos_i' \cdot \hat{\wavevec}_j)\sin(\wavevec_j\cdot\atpos_i')\right]\sin(\wavevec_j\cdot\pos+\displphase_j) \\
    &= \den_0 + \sum_{j=1}^3 \fitden_j \cos(\wavevec_j\cdot\pos+\displphase_j+\pi/2) ,
    \label{eq:den_Acos}
\end{align}
where 
\begin{equation}
    \fitden_j = \frac{\displamp^{(j)}}{\smearing^2} \hat{\wavevec}_j \cdot \sum_{i=1}^N \atpos_i' \exp\left(-\frac{1}{2\smearing^2}(\atpos_i')^2\right)\sin(\wavevec_j\cdot\atpos_i').
    \label{eq:ampdispl2ampden}
\end{equation}
We note the sum in Equation~\ref{eq:ampdispl2ampden} is simply a constant after the summation is performed. This constant depends on the high-symmetry structure and the smearing used. 

Equation~\ref{eq:den_Acos} shows that the smeared atomic density of the structure with a charge density wave is also a Fourier series generated by the three wavevectors. The phases of each component ($\phase_j$) simply differ from those generating the charge density wave displacements ($\displphase_j$) by $\pi/2$. This is the relation between the phases shown in Tables~\ref{tab:displphi_type} and~\ref{tab:phi_type} Importantly, Equation~\ref{eq:ampdispl2ampden} shows that the amplitude of each component $\fitden_j$ is directly proportional to the displacement amplitudes $\displamp^{(j)}$. Therefore, the smeared atomic density associated with a structure with a charge density wave is indeed a Fourier series with the same wavevectors. The above derivation can be generalized to an inhomogeneous charge density wave where both $\displamp^{(j)}$ and $\phase_j$ are functions of space. This is because the arguments or approximations in the derivation do not depend on the spatial dependence of these quantities.

\begin{center}
\begin{table}[htbp!]
    \centering
    \begin{tabular}{|c|c|c|c|}
        \hline
        CDW type & Hollow-center & Filled-center & Hexagonal   \\
        \hline
        $\displphase_1^0\ (2\pi/3)$ & $-\frac{1}{3}$ & $-\frac{1}{3}$ & $-\frac{1}{3}$ \\
        \hline 
        $\displphase_2^0\ (2\pi/3)$ & $\frac{2}{3}$ & $-\frac{1}{3}$ & $-\frac{1}{3}$ \\ 
        \hline
        $\displphase_3^0\ (2\pi/3)$ & $-\frac{1}{3}$ & $\frac{5}{3}$ & $-\frac{1}{3}$ \\ 
        \hline
    \end{tabular}
    \caption{The reference values of the phases that generates the three types of normal state $3\times3$ charge density waves by Equation~\ref{eq:cdw_displ}. }
    \label{tab:displphi_type}
\end{table}
\end{center}

The total energies and relative energies between these three types of CDWs are shown below: 
\begin{table}[htbp!]
    \centering
    \begin{tabular}{|c|c|c|}
    \hline
    Type of CDW  &  Total energy (eV)  &  Relative energy from the most stable CDW (eV) \\ 
    \hline
      Hollow-center CDW  &  -6564.928  &  0  \\
    \hline
      Filled-center CDW  &  -6564.906  &  0.022  \\
    \hline
      Hexagonal CDW  &  -6564.898  &  0.030  \\
    \hline
    \end{tabular}
    \caption{The total energies and relative energies of each type of CDW. For these calculations, $3\times3$ unit cells are included in the simulation cell. The most stable CDW is the hollow-center CDW in our calculations.}
    \label{tab:cdw_enr}
\end{table}

\section{Extracting the smeared atomic density of charge density waves in the MoSe$_2$ layer}
\label{sec:cdwmose2}
We have found that the charge density waves formed in the MoSe$_2$ layer to be very weak, that the background contribution to the oscillations in the smeared atomic density due to moir\'e relaxation becomes comparable to that due to the charge density waves. In order to remove this background, we run another relaxation with an electronic temperature of 5000 K. At this electronic temperature, charge density waves do not develop. We have checked that the moir\'e relaxation pattern is similar to that at 15.7 K. This high-temperature-relaxed structure gives the atomic displacement from the high-symmetry positions due to moir\'e relaxation only. These displacements are subtracted out from the low-temperature-relaxed structures, so that a structure with only charge density wave relaxations is obtained. The smeared atomic density is then calculated based on this structure. 

\section{Estimating the phase field with the local phase approximation}
\label{sec:lpa_derivation}

We prove Equation~\ref{eq:phi_tan} for a uniform charge density wave, where the phases $(\phase_i)$, amplitudes $(\den_i)$ of the $i$-th components, and the mean $(\den_0)$ are constants. In this case, the Fourier series for the smeared atomic density in Equation~\ref{eq:rho_ft} takes the form
\begin{equation}
    \fitden(\pos) = \fitden_0 + \sum_{i=1}^{3}\fitden_{i}\cos(\wavevec_i\cdot\pos+\phase_i). 
    \label{eq:den_lpa}
\end{equation}

By taking gradients of both sides, one obtains
\begin{equation}
    \nabla\fitden(\pos) = - \sum_{i=1}^{3}\fitden_i \wavevec_i\sin(\wavevec_i\cdot\pos+\phase_i).
    \label{eq:graddenft}
\end{equation}

If one projects both sides along the unit normal vector $\hat{\normvec}_1$ to the wavevector $\wavevec_1$, for instance, then the $i=1$ component in Equation~\ref{eq:graddenft} vanishes since $\hat{\normvec}_1\cdot\wavevec_1=0$. So Equation~\ref{eq:graddenft} becomes
\begin{equation}
    \hat{\normvec}_1\cdot\nabla\fitden(\pos) = - \sum_{i=2}^{3}\fitden_i (\hat{\normvec}_1\cdot\wavevec_i)\cos(\wavevec_i\cdot\pos+\phase_i).
    \label{eq:ngraddenft}
\end{equation}
Now if one takes the gradient on both sides again but projects both sides along the normal vector $\hat{\normvec}_2$ to the wavevector $\wavevec_2$ instead, then the $i=2$ component also vanishes. This leaves one with the expression
\begin{equation}
    \hat{\normvec}_2\cdot\nabla(\hat{\normvec}_1\cdot\nabla\fitden(\pos)) = \fitden_3 (\hat{\normvec}_2\cdot\wavevec_3)(\hat{\normvec}_1\cdot\wavevec_3)\sin(\wavevec_3\cdot\pos+\phase_3).
    \label{eq:denft_3}
\end{equation}

Equation~\ref{eq:denft_3} allows one to calculate $\phase_3$ by taking the $\arcsin$ on both sides. Geometrically, it shows that the projected curvature along the directions $\normvec_1$ and $\normvec_2$ of a three-component Fourier series is determined by the oscillation along the remaining wavevector $\wavevec_3$ only. This is what one would expect intuitively. If one picks a position $\pos_0$, and consider gradient of $\fitden$ along $\normvec_1$, then the contribution to the gradient due to the Fourier component of $\wavevec_1$ is locally zero. This is because the gradient is being calculated along a plane of constant phase of the $\wavevec_1$-component. If one subsequently considers the variation of this gradient along the direction $\normvec_2$, the variation of the gradient generated by the Fourier components of $\wavevec_{1,2}$, i.e. the projected curvature, are both locally zero. This is because the projected curvature is being calculated along the planes of constant phases of the Fourier components of $\wavevec_{1,2}$. The only contribution to the projected curvature is therefore associated with the Fourier component of $\wavevec_3$ only. 

Solving Equation~\ref{eq:denft_3} at any chosen position $\pos_0$ gives two values of $\phase_3$. In order to determine $\phase_3$ unambiguously, one needs to know the value of $\cos(\wavevec_3\cdot\pos_0+\phi_3)$ as well. This is done by taking the gradient of Equation~\ref{eq:denft_3}, which gives
\begin{equation}
    \nabla(\hat{\normvec}_2\cdot\nabla(\hat{\normvec}_1\cdot\nabla\fitden(\pos))) = - \fitden_3 \wavevec_3 (\hat{\normvec}_2\cdot\wavevec_3)(\hat{\normvec}_1\cdot\wavevec_3)\cos(\wavevec_3\cdot\pos+\phase_3).
    \label{eq:graddenft_3}
\end{equation}
The value of $\cos(\wavevec_3\cdot\pos+\phase_3)$ can be solved from Equation~\ref{eq:graddenft_3} if it is a scalar equation. Therefore, we project both sides along the wavevector $\wavevec_3$ to obtain
\begin{equation}
    \wavevec_3\cdot\nabla(\hat{\normvec}_2\cdot\nabla(\hat{\normvec}_1\cdot\nabla\fitden(\pos))) = - \fitden_3 |\wavevec_3|^2 (\hat{\normvec}_2\cdot\wavevec_3)(\hat{\normvec}_1\cdot\wavevec_3)\cos(\wavevec_3\cdot\pos+\phase_3).
    \label{eq:qgraddenft_3}
\end{equation}
Now, if we divide Equation~\ref{eq:denft_3} by Equation~\ref{eq:qgraddenft_3}, the factor $\fitden_3$ cancels out, and one obtains Equation~\ref{eq:phi_tan}. Knowing the values of both $\cos(\wavevec_3\cdot\pos+\phase_3)$ and $\sin(\wavevec_3\cdot\pos+\phase_3)$, we unambiguously determine the value of $\wavevec_3\cdot\pos+\phase_3$ using the \textit{numpy} function \textit{numpy.arctan2}. We then subtract out $\wavevec_3\cdot\pos$ to obtain the value of $\phase_3$. We note that the value of $\phase_3$ therefore obviously depends on the choice of origin. The other phases $\phase_1$ and $\phase_2$ can be obtained similarly by changing the normal vectors of projection to $\normvec_2$ and $\normvec_3$, and $\normvec_3$ and $\normvec_1$ respectively.

The values of the phase of the density oscillation associated with each type of CDW are summarized in Table~\ref{tab:phi_type}. We note that there are infinitely many valid choices of phases for a given type of charge density wave. Intuitively, for instance, if a filled-center charge density is shifted to a different Se atom, the values of the phases will obviously change but the type of charge density wave remains the same. From this consideration, it is possible to show that the charge density wave generated by the phase triple $\{\phase_1^0,\phase_2^0,\phase_3^0\}$ can also be generated by the family of phases, 
\begin{align}
    \phase_1(\pos_p) &= \phase_1^{0} + \frac{2\pi}{3} M \\
    \phase_2(\pos_p) &= \phase_2^{0} + \frac{2\pi}{3} N \\ 
    \phase_3(\pos_p) &= \phase_3^{0} - \frac{2\pi}{3} (M+N) ,
\end{align}
where $N,M$ are integers.

\newlength{\tabgap}
\setlength{\tabgap}{4pt}
\begin{center}
\begin{table}[htbp!]
    \centering
    \begin{tabular}{|c|c|c|c|}
        \hline
        CDW type & Hollow-center & Filled-center & Hexagonal   \\
        \hline
        $\phase_1^0\ (2\pi/3)$ & $\frac{23}{12}$ & $\frac{23}{12}$ & $\frac{23}{12}$ \\
        \hline 
        $\phase_2^0\ (2\pi/3)$ & $\frac{35}{12}$ & $\frac{23}{12}$ & $\frac{23}{12}$ \\ 
        \hline
        $\phase_3^0\ (2\pi/3)$ & $\frac{23}{12}$ & $\frac{11}{12}$ & $\frac{23}{12}$ \\ 
        \hline
    \end{tabular}
    \caption{The reference values of the phases that generates the three types of normal state $3\times3$ charge density waves using Equation~\ref{eq:rho_ft_const}. }
    \label{tab:phi_type}
\end{table}
\end{center}

For approximating an inhomogeneous charge density wave with Equation~\ref{eq:rho_ft}, we assume that the phase is locally constant so that Equation~\ref{eq:den_lpa} is valid at each estimation point. This is equivalent to ignoring the terms first-order or higher-order in the gradient of the phase, $\nabla\phase_i$. 

We have found that the local phase approximation works even for cases where the phase is linear in space, i.e. $\phase_i(\pos)=-\wavevec_i\cdot\pos$. This phase field describes a charge density wave where the $i$-th component vanishes. We found that the phase profile estimated using this method reproduces the linear phase profile. Furthermore, the amplitude of the vanishing $i$-th component is correctly calculated to be zero using the estimated phase profile. This shows that the local phase approximation is appropriate for our purpose of estimating the amplitude of the charge density waves.   

\section{List of derivatives of the smeared atomic density}
\label{sec:derivsden}
In order to analytically calculate the value of the phases, we rewrite Equations~\ref{eq:denft_3} and \ref{eq:graddenft_3} by replacing the fitting density $\fitden(\pos)$ with the exact smeared atomic density $\den(\pos)$. This leads to the appearance of the derivatives of the Gaussians. Below, we list the expressions used for evaluating the phases analytically for reference. We note only the atomic positions and the smearing need to be known for computing all quantities. 

The gradient of the exact smeared atomic density can be written as
\begin{equation}
    \nabla\den(\pos)=\ihat \partial_x\den + \jhat \partial_y \den. 
    \label{eq:gradden_comp}
\end{equation}
The dot product between the $k$-th normal vector $\normvec_k$ and the gradient of the density relevant for Equation~\ref{eq:ngraddenft} can then be written as
\begin{equation}
    \hat{\normvec}_k\cdot\nabla\den(\pos) = \hat{\nnormvec}_{kx} \partial_x\den + \hat{\nnormvec}_{ky} \partial_y \den.
    \label{eq:ngradden_comp}
\end{equation}
The projected curvature along the normal vectors $\normvec_{j,k}$ relevant for Equation~\ref{eq:denft_3} can be written as 
\begin{equation}
    \hat{\normvec}_j\cdot\nabla(\hat{\normvec}_k\cdot\nabla\den(\pos)) = \hat{\nnormvec}_{jx} \hat{\nnormvec}_{kx} \partial^2_{xx}\den + ( \hat{\nnormvec}_{jy} \hat{\nnormvec}_{kx} + \hat{\nnormvec}_{jx} \hat{\nnormvec}_{ky} ) \partial^2_{xy} \den + \hat{\nnormvec}_{jy} \hat{\nnormvec}_{ky} \partial^2_{yy}\den .
    \label{eq:ngradngradden_comp}
\end{equation}
The gradient of the projected curvature relevant for Equation~\ref{eq:graddenft_3} can be written as
\begin{multline}
    \nabla(\hat{\normvec}_j\cdot\nabla(\hat{\normvec}_k\cdot\nabla\den(\pos))) = \ihat [\hat{\nnormvec}_{jx} \hat{\nnormvec}_{kx} \partial^3_{xxx}\den +
    \hat{\nnormvec}_{jy} \hat{\nnormvec}_{ky} \partial^3_{xyy}\den + 
    (\hat{\nnormvec}_{jx} \hat{\nnormvec}_{ky} + \hat{\nnormvec}_{jy} \hat{\nnormvec}_{kx} ) \partial^3_{xxy}\den]  \\
    + \jhat [\hat{\nnormvec}_{jy} \hat{\nnormvec}_{ky} \partial^3_{yyy} \den + 
    \hat{\nnormvec}_{jx} \hat{\nnormvec}_{kx} \partial^3_{yxx} \den + 
    (\hat{\nnormvec}_{jx} \hat{\nnormvec}_{ky}  + \hat{\nnormvec}_{jy} \hat{\nnormvec}_{kx}) \partial^3_{yxy}\den ].
    \label{eq:gradngradngradden_comp}
\end{multline}
Thus the dot product between the $\wavevec_i$ and the gradient of the projected curvature, which is relevant for Equation~\ref{eq:qgraddenft_3}, is
\begin{multline}
    \wavevec_i\cdot\nabla(\hat{\normvec}_j\cdot\nabla(\hat{\normvec}_k\cdot\nabla\den(\pos))) = \wwavevec_{ix} [\hat{\nnormvec}_{jx} \hat{\nnormvec}_{kx} \partial^3_{xxx}\den +
    \hat{\nnormvec}_{jy} \hat{\nnormvec}_{ky} \partial^3_{xyy}\den + 
    (\hat{\nnormvec}_{jx} \hat{\nnormvec}_{ky} + \hat{\nnormvec}_{jy} \hat{\nnormvec}_{kx} ) \partial^3_{yxx}\den]  \\
    + \wwavevec_{iy} [\hat{\nnormvec}_{jy} \hat{\nnormvec}_{ky} \partial^3_{yyy} \den + 
    \hat{\nnormvec}_{jx} \hat{\nnormvec}_{kx} \partial^3_{yxx} \den + 
    (\hat{\nnormvec}_{jx} \hat{\nnormvec}_{ky}  + \hat{\nnormvec}_{jy} \hat{\nnormvec}_{kx}) \partial^3_{yxy}\den ].
    \label{eq:qgradngradngradden_comp}
\end{multline}

Inspecting Equations~\ref{eq:gradden_comp} to \ref{eq:gradngradngradden_comp}, the list of derivatives needed for computation is
\begin{equation}
    \partial_x\den = -(2\pi\smearing^{4})^{-1}\sum_{i=1}^{N}(x-X_i)\denexp,
\end{equation}
\begin{equation}
    \partial_y\den = -(2\pi\smearing^{4})^{-1}\sum_{i=1}^{N}(y-Y_i)\denexp,
\end{equation}
\begin{equation}
    \partial^2_{xx}\den = -\smearing^{-2}\den + (2\pi\smearing^{6})^{-1}\sum_{i=1}^{N}(x-X_i)^2\denexp,
\end{equation}
\begin{equation}
    \partial^2_{xy}\den = (2\pi\smearing^{6})^{-1}\sum_{i=1}^{N}(x-X_i)(y-Y_i)\denexp,
\end{equation}
\begin{equation}
    \partial^2_{yy}\den = -\smearing^{-2}\den +(2\pi\smearing^{6})^{-1}\sum_{i=1}^{N}(y-Y_i)^2\denexp,
\end{equation}
\begin{equation}
    \partial^3_{xxx}\den = -3\smearing^{-2}\partial_x\den + (2\pi\smearing^{8})^{-1} \sum_{i=1}^{N}(x-X_i)^3\denexp,
\end{equation}
\begin{equation}
    \partial^3_{yxx}\den = -\smearing^{-2}\partial_y\den + (2\pi\smearing^{8})^{-1} \sum_{i=1}^{N}(x-X_i)^2(y-Y_i)\denexp,
\end{equation}
\begin{equation}
    \partial^3_{yxy}\den = -\smearing^{-2}\partial_x\den + (2\pi\smearing^{8})^{-1} \sum_{i=1}^{N}(x-X_i)(y-Y_i)^2\denexp,
\end{equation}
\begin{equation}
    \partial^3_{yyy}\den = -3\smearing^{-2}\partial_y\den + (2\pi\smearing^{8})^{-1} \sum_{i=1}^{N}(y-Y_i)^3\denexp.
\end{equation}

\section{Euler-Lagrange equations for local amplitude and mean estimation}
\label{sec:el_amp}
In order to obtain the local amplitudes of each component and the local mean density at a fitting $\pos_p$, the Lagrangian defined in Equation~\ref{eq:amp_L} is used for numerical optimization. We use the gradient descent method to derive the equations for numerically obtaining $\fitden_0(\pos_p)$ and $\fitden_{i}(\pos_p)$, which are 
\begin{align}
    \fitden_0^{(n+1)}(\pos_p) &= \fitden_0^{(n)}(\pos_p) - \gamma \left(\frac{\partial{\Lagr}}{\partial{\fitden_0}}\right)^{(n)} \\
    \fitden_{i}^{(n+1)}(\pos_p) &= \fitden_{i}^{(n)}(\pos_p) - \gamma \left(\frac{\partial{\Lagr}}{\partial{\fitden_i}}\right)^{(n)},
    \label{eq:amp_opt}
\end{align}
where $\gamma$ is the step length, $\fitden_0^{(n)}(\pos_p)$ and $\fitden_{i}^{(n)}(\pos_p)$ are the $n$-th iterates for the estimated local mean $\fitden_0(\pos_p)$ and the local amplitudes $\fitden_{i}(\pos_p)$ respectively, and $\left(\frac{\partial{\Lagr}}{\partial{\fitden_i}}\right)^{(n)}$ are the derivatives of the Lagrangian computed using Equation~\ref{eq:amp_L} at the $n$-th iteration. The step length is taken to be 0.1 in our calculations. The fitting area $A_p$ is taken to be a parallelogram centered at the fitting point $\pos_p$ and spanned by the lattice vectors $3\vc{a}_1$ and $3\vc{a}_2$. 

The local mean density $\fitden_0(\pos_p)$ can then be shown to be given by
\begin{equation}
    \fitden_0(\pos_p)=\frac{1}{A_p}\int_{A_p}\den(\pos)d^2\pos,
    \label{eq:mean}
\end{equation}
which is what one will expect intuitively. Therefore, no iteration is required for estimating the local mean density.

For propagating the iterates of the local amplitude of the $i$-th component $\fitden_{i}(\pos_p)$, the partial derivative of $\Lagr^{(n)}$ against $\fitden_{i}$ is computed using 
\begin{equation}
    \left(\frac{\partial\Lagr}{\partial\fitden_{i}}\right)^{(n)}=-2\int_{A_p}\penal^{(n)}(\pos)\sin(\wavevec_i\cdot\pos+\phase_i(\pos))d^2\pos,
    \label{eq:amp_deriv}
\end{equation}
where $\phase_i(\pos)$ is the phase field computing using Equation~\ref{eq:phi_tan}, and $\penal^{(n)}(\pos)$ is the penalty function computed at the $n$-th iteration. The penalty function is defined as
\begin{equation}
    \penal^{(n)}(\pos) = \fitden^{(n)}(\pos) - \den^{(n)}(\pos).
    \label{eq:penal}
\end{equation}

The convergence condition for $\fitden_{i}$ is when the root-mean-square error $\varepsilon^{(n)}$, defined as,
\begin{equation}
    \varepsilon^{(n)} = \sqrt{\frac{1}{A_p}\int_{A_p} [\penal^{(n)}(\pos)]^2 d^2\pos},
\end{equation}
is smaller than some threshold. The threshold is taken as $10^{-3}~\text{atom}\Ang^{-2}$ for the CDW of the Nb atoms, and $2\times10^{-5}~\text{atom}\Ang^{-2}$ for that of the Mo atoms. 

The displacement amplitude is related to the density oscillation amplitude by $\displamp_j = 117~\Ang^3\mathrm{atom}^{-1} \times \fitden_j$ for $\smearing=0.85\lattconst$. The constant of proportionality obtained by first generating a set of CDWs with different displacement amplitudes, followed by calculating the atomic density amplitude of these structures. The slope of the variation of the displacement amplitude against the atomic density amplitude (averaged over the three components) is then calculated. This variation is shown in Supplementary Fig.~\ref{fig:displ_rho_fit}.

\begin{figure}[htbp!]
    \centering
    \includegraphics[width=0.75\linewidth]{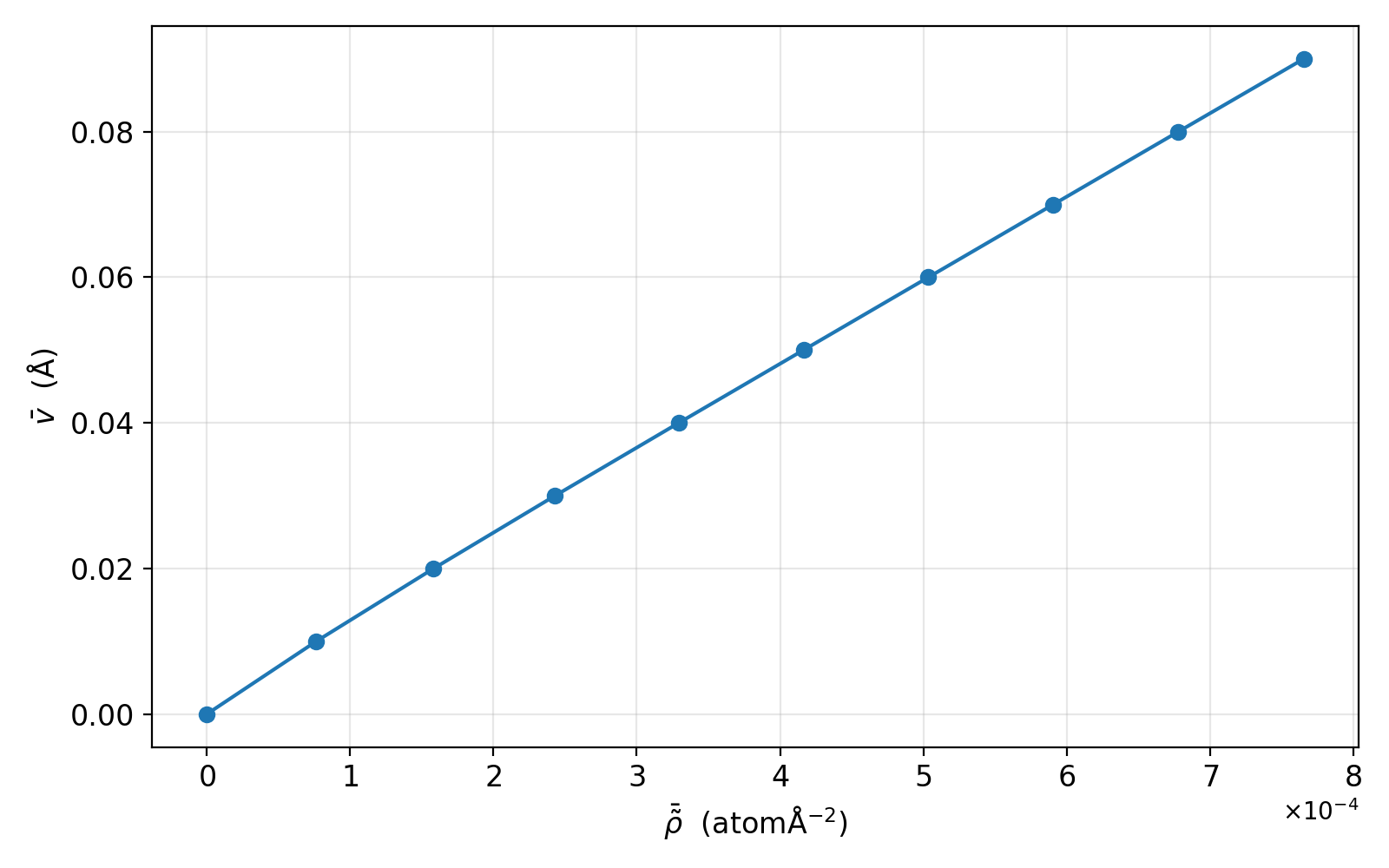}
    \caption{The variation of the displacement amplitude, $\bar{\displamp}$ against the amplitude of the atomic density $\bar{\fitden}$.}
    \label{fig:displ_rho_fit}
\end{figure}

\begin{figure}[htbp!]
    \centering
    \includegraphics[width=0.5\linewidth]{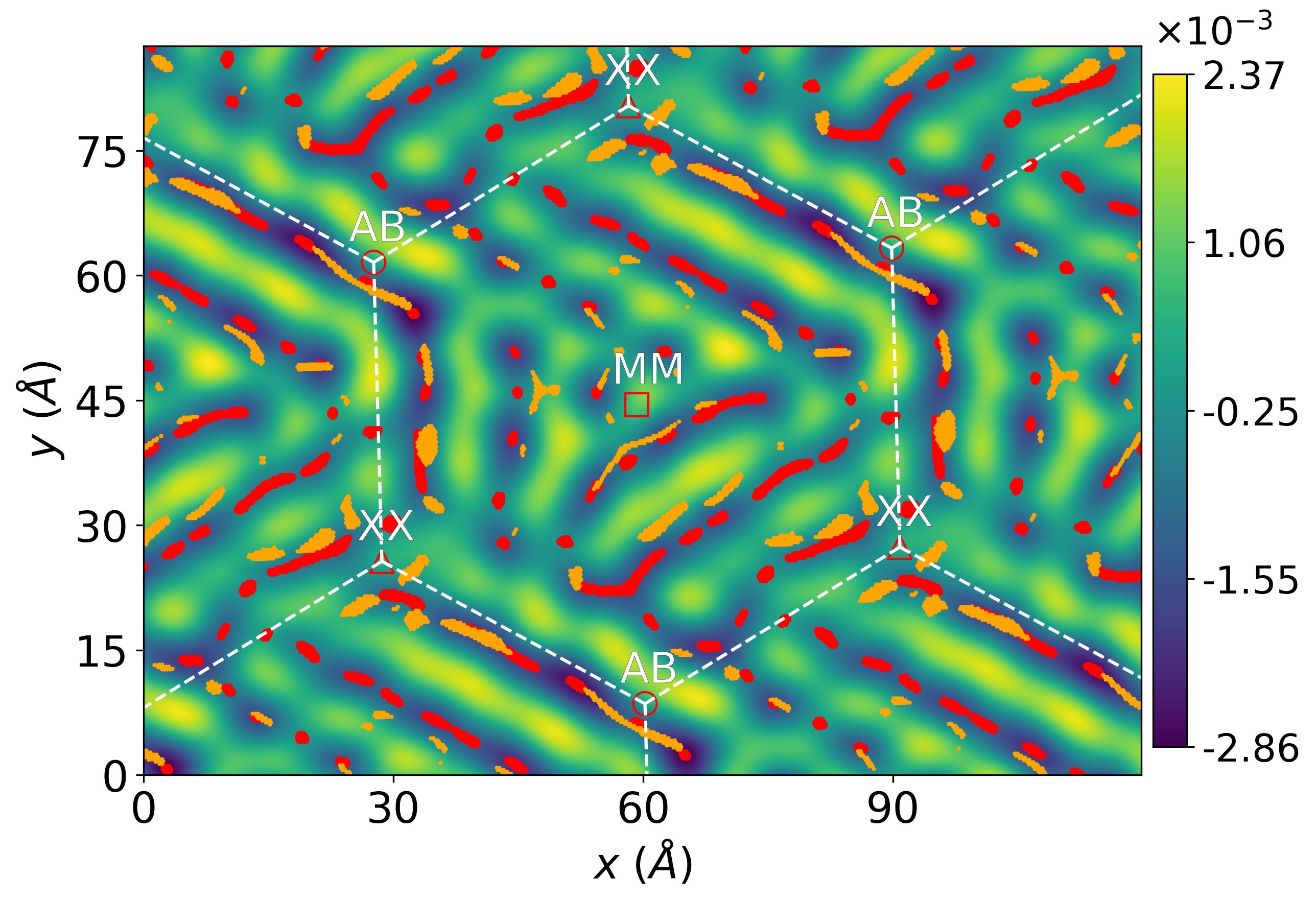}
    \caption{Comparison of the positions of the extrema in the smeared Mo density to the smeared Nb density of the structure reported in the main text. The maxima (minima) in the smeared Mo density are marked in orange (red), and the smeared Nb density is plotted in the background. In these calculations, $\smearing=0.85\lattconst$.}
    \label{fig:compare_extr_den}
\end{figure}

\newcommand{\figfac}{0.55}
\begin{figure*}[htbp!]
    \centering
    \begin{subfigure}[h]{\figfac\textwidth}
        \centering
        \caption{}
        \includegraphics[width=\textwidth]{kbm/MM/displ.png}
    \end{subfigure}%
    \par
    \begin{subfigure}[h]{\figfac\textwidth}
        \centering
        \caption{}
        \includegraphics[width=\textwidth]{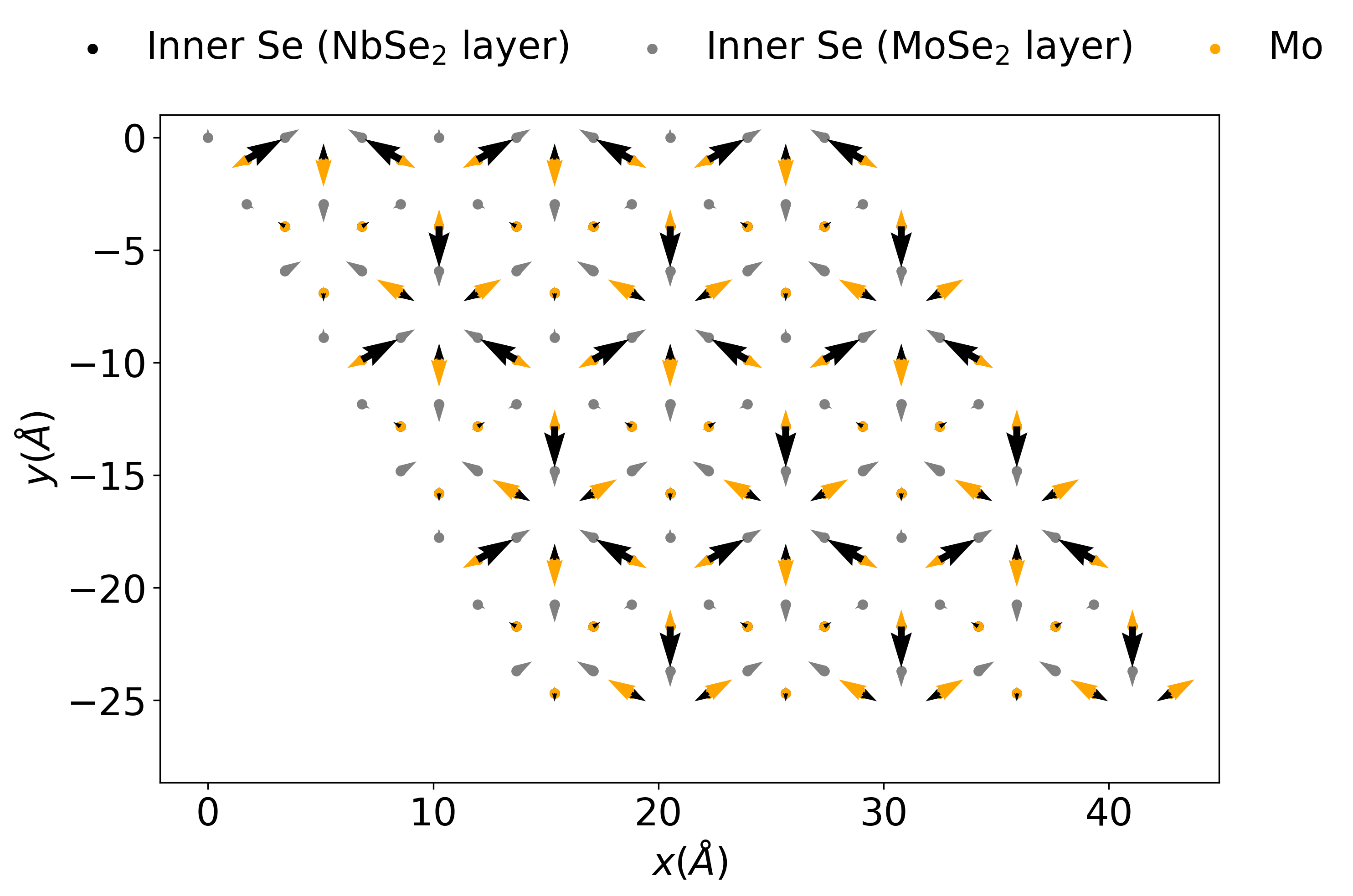}
    \end{subfigure}%
    \par
    \begin{subfigure}[h]{\figfac\textwidth}
        \centering
        \caption{}
        \includegraphics[width=\textwidth]{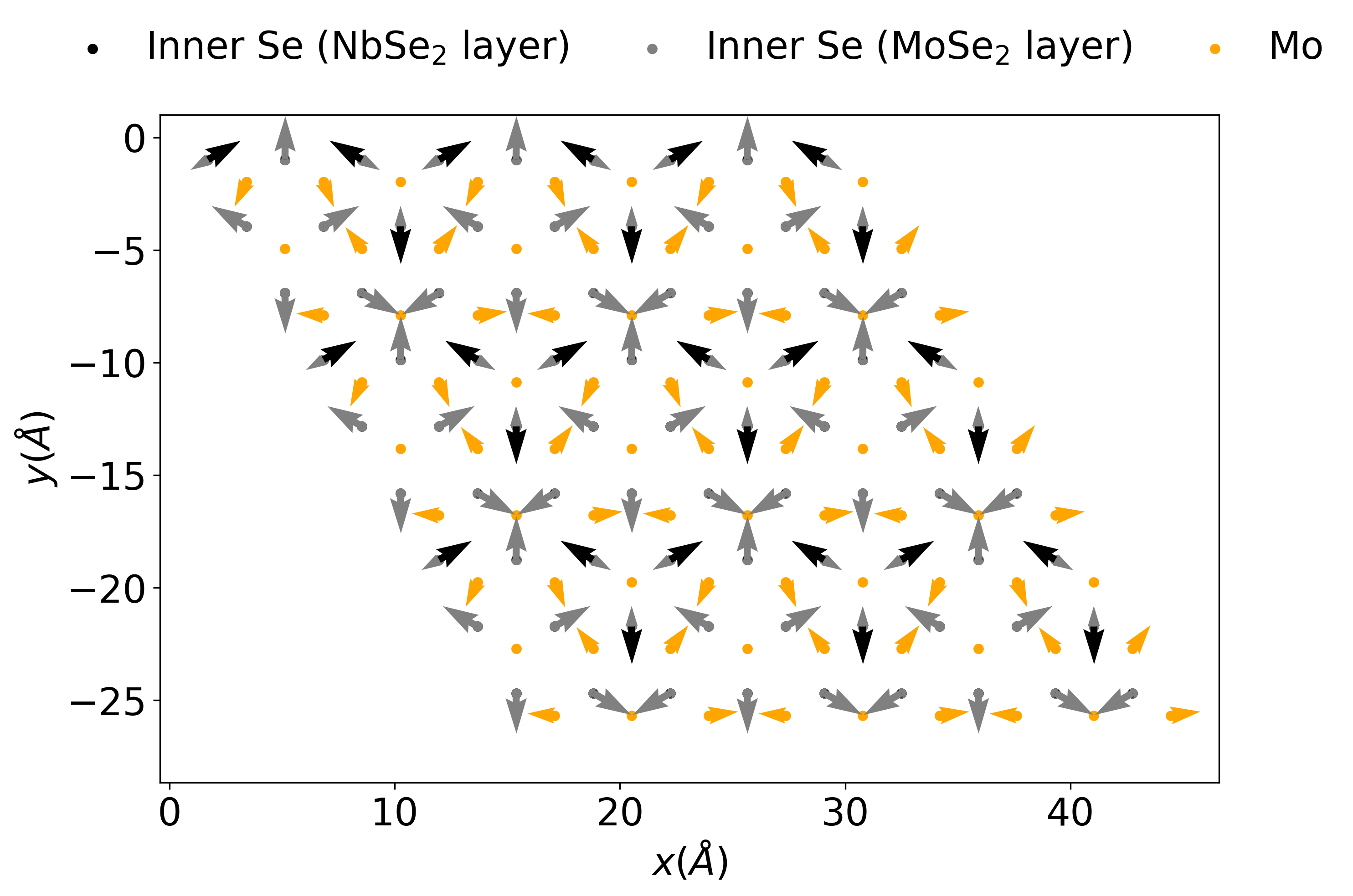}
    \end{subfigure}%
    \caption{The atomic displacements of the inner Se atoms in the NbSe$_2$ layer (black), MoSe$_2$ layer (grey), and Mo atoms (orange) in the untwisted (a): MM-, (b): AB-, (c): XX-stacked bilayers of NbSe$_2$/MoSe$_2$.}
    \label{fig:cdw_displ_highsym_all}
\end{figure*}

\clearpage
\section{Metastable relaxed twisted bilayer structure}
\label{sec:metastable_struct}

We note that more symmetrical metastable structures with somewhat different amplitude variations have been found. The total energy of the most stable structure reported in the main text is -1044573.2080~eV, while that of the metastable structure reported here is -1044573.0424~eV. The metastable structure is less stable by 0.2~eV, which is equivalent to 0.3~meV per unit cell. 

\begin{figure*}[htbp!]
    \begin{subfigure}[t]{0.49\textwidth}
        \centering
        \caption{\hspace{\captionhspacea}NbSe$_2$ layer}
        \includegraphics[width=\textwidth]{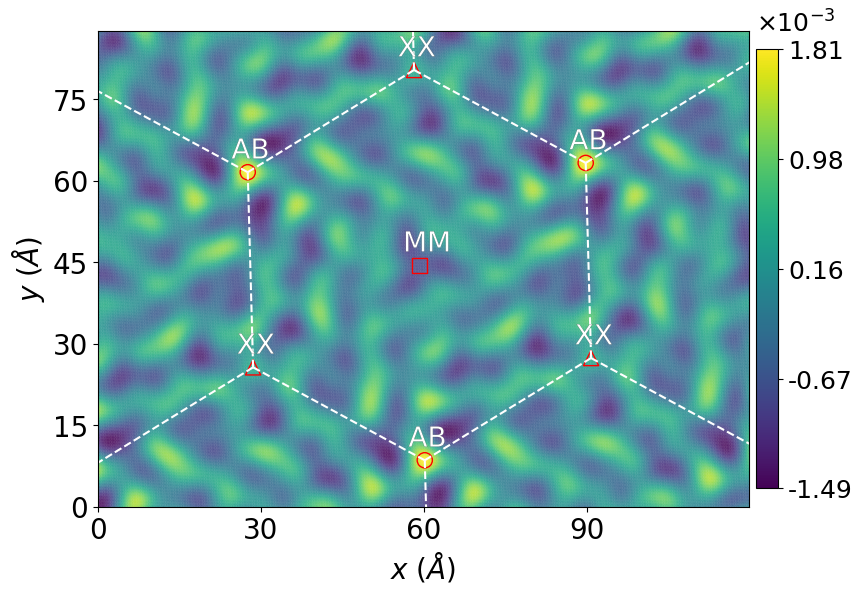}
    \end{subfigure}
    ~  
    \begin{subfigure}[t]{0.49\textwidth}
        \centering
        \caption{\hspace{\captionhspacea}MoSe$_2$ layer}
        \includegraphics[width=\textwidth]{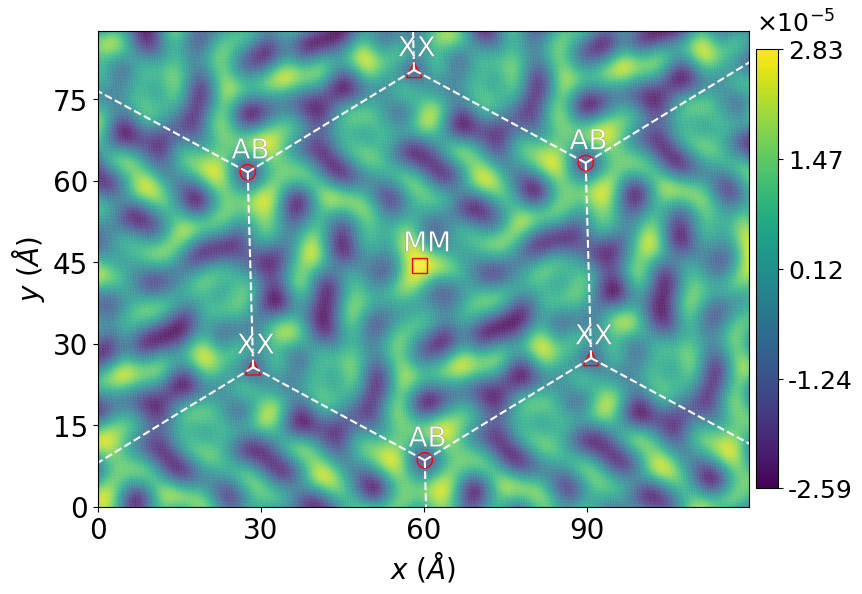}
    \end{subfigure}
    \caption{The (a): smeared Nb atom density, and (b): smeared Mo atom density computed using $\smearing=0.85\lattconst$. The units are atoms$\Ang^{-2}$.}
    \label{fig:meta_den_tb}
\end{figure*}

Supplementary Figures~\ref{fig:meta_den_tb}(a) and~(b) show the smeared atomic density of the Nb and Mo atoms in the metastable twisted bilayer NbSe$_2$/MoSe$_2$. This metastable structure has a different CDW domain distribution that demonstrates threefold rotation symmetry around all high-symmetry centers. See Supplementary Figures~\ref{fig:meta_typeamp_tb}(a) and~(b). However, the observation that the density minima in the smeared Nb density overlaps with the density maxima in the smeared Mo density still holds, see Supplementary Fig.~\ref{fig:meta_optden}.  

The CDW displacement amplitudes in the NbSe$_2$ and MoSe$_2$ layers are shown in Supplementary Figures~\ref{fig:meta_typeamp_tb}(c) and~(d). The local CDW displacement amplitudes in the NbSe$_2$ are $2.9\times10^{-2}~\Ang$ at the MM center; $5.9\times10^{-2}~\Ang$ at the AB center; $3.0\times10^{-2}~\Ang$ at the XX center. The local CDW displacement amplitudes in the MoSe$_2$ layer are $7.2\times10^{-4}~\Ang$ at the MM center; $8.4\times10^{-4}~\Ang$ at the AB center; and $5.9\times10^{-4}~\Ang$ at the XX center. In the MoSe$_2$ layer, the amplitudes and the AB and MM stacking regions have the same order of magnitude but are reduced compared to the most stable structure, and the amplitude becomes larger at the AB center than at the MM center, see Supplementary Fig.~\ref{fig:meta_typeamp_tb}. However, the displacement amplitude remains the smallest at the XX center. The predicted displacement amplitudes of the CDW in the MoSe$_2$ layer based on Fig.~\ref{fig:amp_amp} are $4.9\times10^{-4}~\Ang$ at the MM center; $8.3\times10^{-4}~\Ang$ at the AB center; and $4.2\times10^{-4}~\Ang$ at the XX center respectively. The values of the CDW displacement amplitudes in the NbSe$_2$ and MoSe$_2$ layers at the AB and XX centers are therefore consistent with that predicted by Fig.~\ref{fig:amp_amp}. The enhancement of CDW displacement amplitude at the MM center is also observed. 

\definecolor{pythoncyan}{HTML}{00FFFF}
\definecolor{pythonorange}{HTML}{FFA500}
\definecolor{pythonpurple}{HTML}{800080}
\begin{figure*}[htbp!]
    \begin{subfigure}[t]{0.49\textwidth}
        \centering
        \caption{\hspace{\captionhspaceb}NbSe$_2$ layer}
        \includegraphics[width=\textwidth]{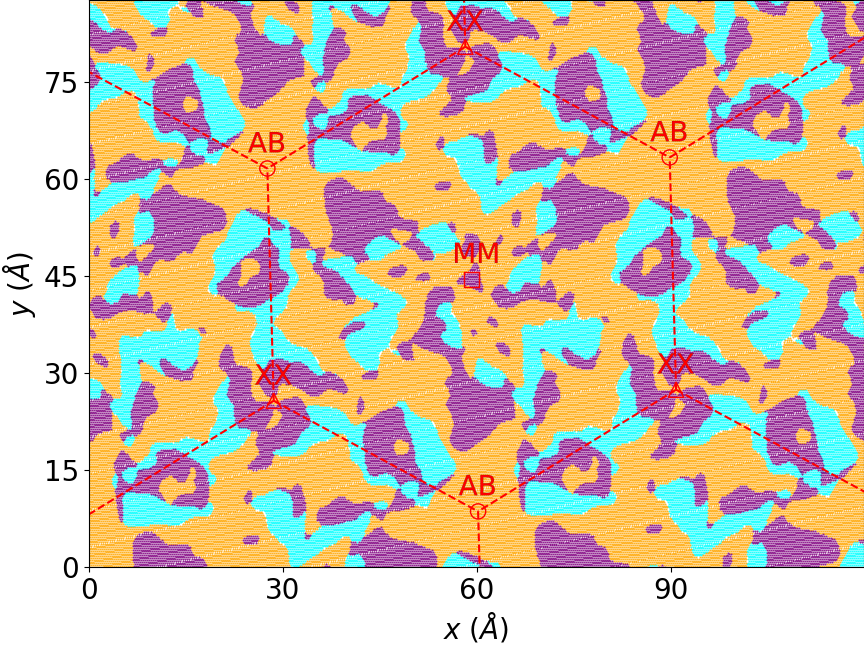}
    \end{subfigure}
    ~  
    \begin{subfigure}[t]{0.49\textwidth}
        \centering
        \caption{\hspace{\captionhspaceb}MoSe$_2$ layer}
        \includegraphics[width=\textwidth]{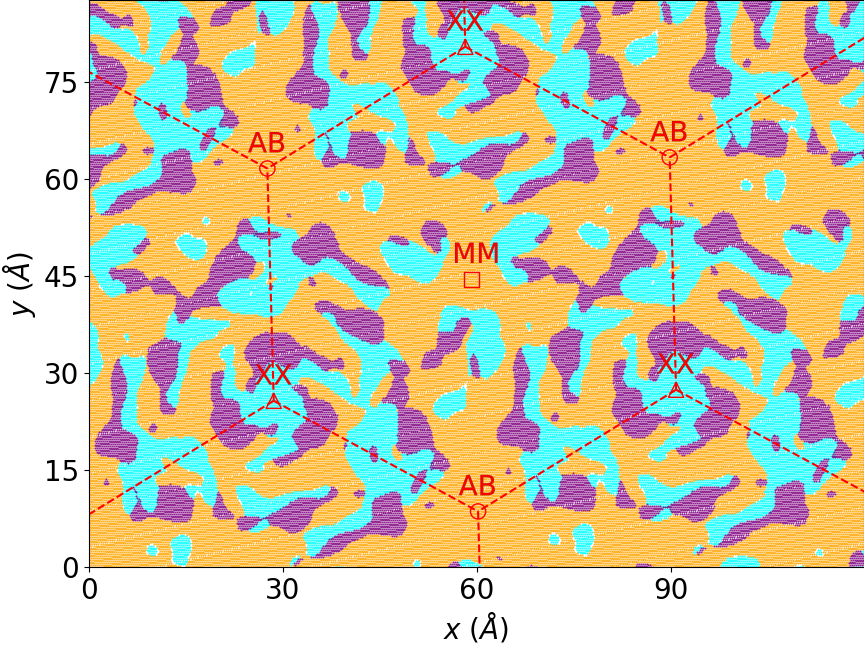}
    \end{subfigure}
    ~  
    \begin{subfigure}[t]{0.49\textwidth}
        \centering
        \caption{\hspace{\captionhspacea}NbSe$_2$ layer}
        \includegraphics[width=\textwidth]{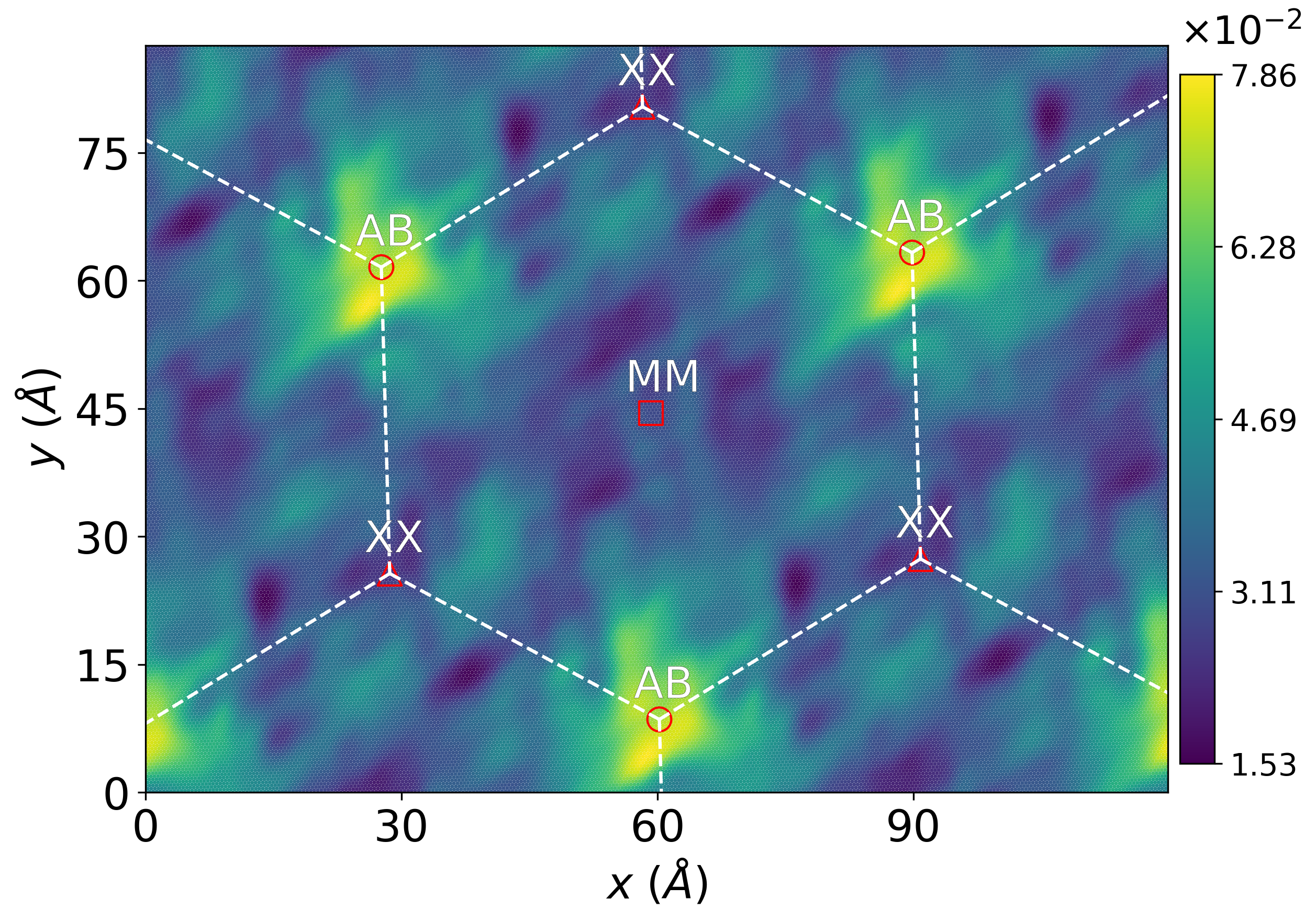}
    \end{subfigure}
    ~  
    \begin{subfigure}[t]{0.49\textwidth}
        \centering
        \caption{\hspace{\captionhspacea}MoSe$_2$ layer}
        \includegraphics[width=\textwidth]{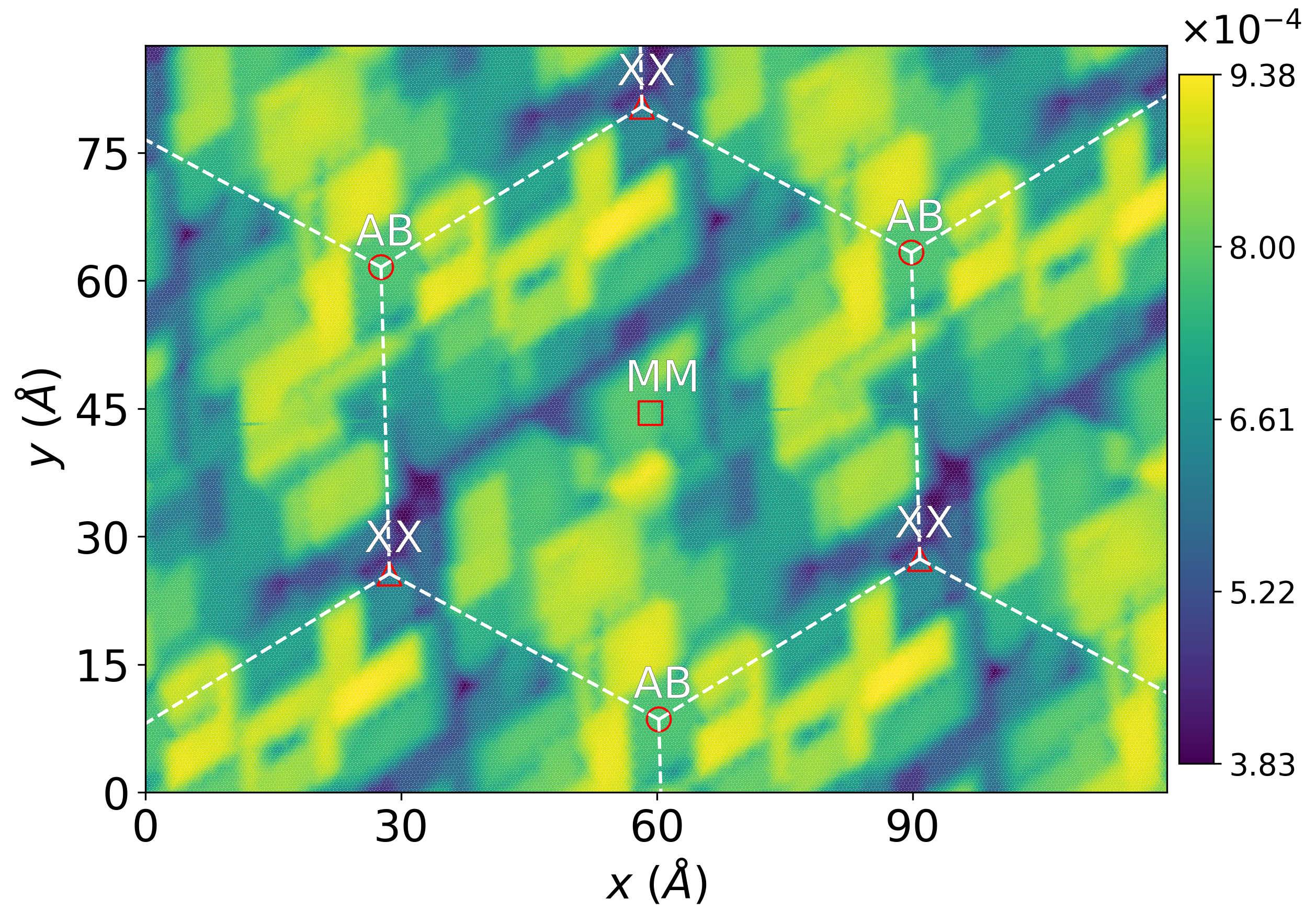}
    \end{subfigure}

    \vspace{10pt}
    Legend:~\colorbox{pythonorange}{Hollow-center CDW}    \colorbox{pythoncyan}{Filled-center CDW} \colorbox{pythonpurple}{\textcolor{white}{Hexagonal CDW}}
    
    \caption{The type identification results in (a): the NbSe$_2$ layer and (b): in the MoSe$_2$ layer. The average CDW displacement amplitude of (c): Nb atoms and (d): Mo atoms in the metastable twisted MoSe$_2$/NbSe$_2$ bilayer. The units are $\Ang$.}
    \label{fig:meta_typeamp_tb}
\end{figure*}

\begin{figure}
    \centering
    \includegraphics[width=0.5\linewidth]{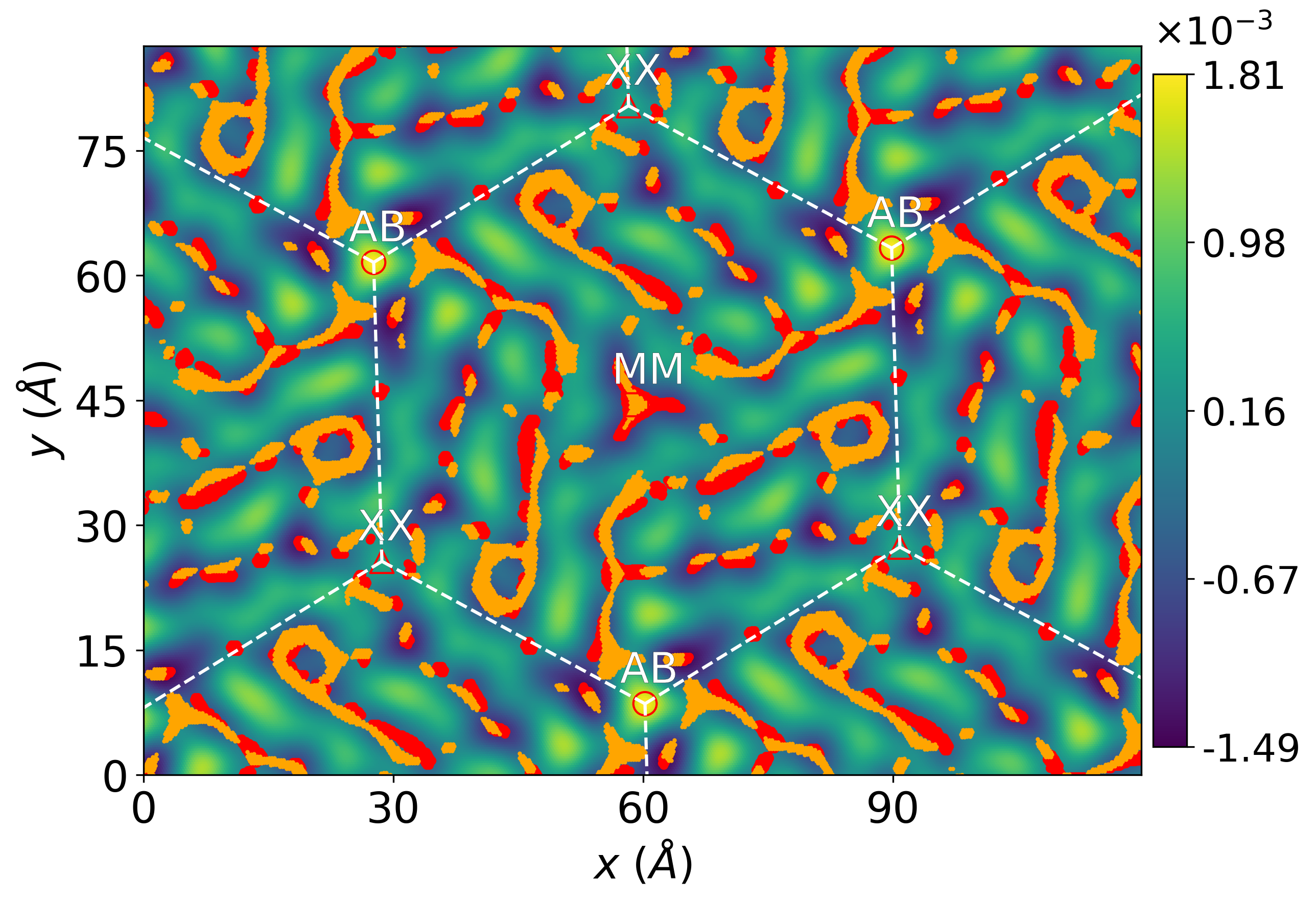}
    \caption{Comparison of the positions of the extrema in the smeared Mo density to the smeared Nb density of the metastable structure. The maxima (minima) in the smeared Mo density are marked in orange (red), and the smeared Nb density is plotted in the background.}
    \label{fig:meta_optden}
\end{figure}

\clearpage
\section{In-plane variation of charge transfer in the NbSe$_2$ layer}
\label{sec:moireqnbse2}
We calculate the net charge on each unit cell in the NbSe$_2$ layer using the Hirshfeld, Voronoi, and Mulliken partition schemes implemented in SIESTA. To obtain the charge distribution, we first perform three interpolations of the charge per atom for the outermost Se, Nb, and innermost Se atoms separately. The interpolated atomic charges for each species are then added to obtain the interpolated charges on each unit cell. 

The variations of the Hirshfeld, Voronoi, and Mulliken charges per unit cell along the diagonal of the moir\'e unit cell (the line traversing the MM-AB-XX-MM centers) are shown in Figures~\ref{fig:q_inplane}(a),~(b), and~(c) respectively. The vertical lines indicate the high-symmetry centers, and the horizontal lines indicate the values of the local charges in the corresponding high-symmetry stacked untwisted bilayer. It can be seen that the net charge fluctuates around the values in the untwisted bilayers with the same stacking except at the MM centers. At the MM centers, the net charge is more negative than the value in the MM-stacked untwisted bilayer. We note that in the Mulliken partition scheme, the number of electrons are reported instead of the atomic net charge. Nonetheless, the number of electrons are the largest at the MM centers. This schemes thus also predicts that the net charge is the more negative at the MM centers than that in the MM-stacked untwisted bilayers. 

\newcommand{\figfacinplct}{0.44}
\newcommand{\hsinplct}{54pt}
\begin{figure*}[htbp!]
    \centering
    \begin{subfigure}[h]{\figfacinplct\textwidth}
        \centering
        \caption{\hspace{\hsinplct} }
        \includegraphics[width=\textwidth]{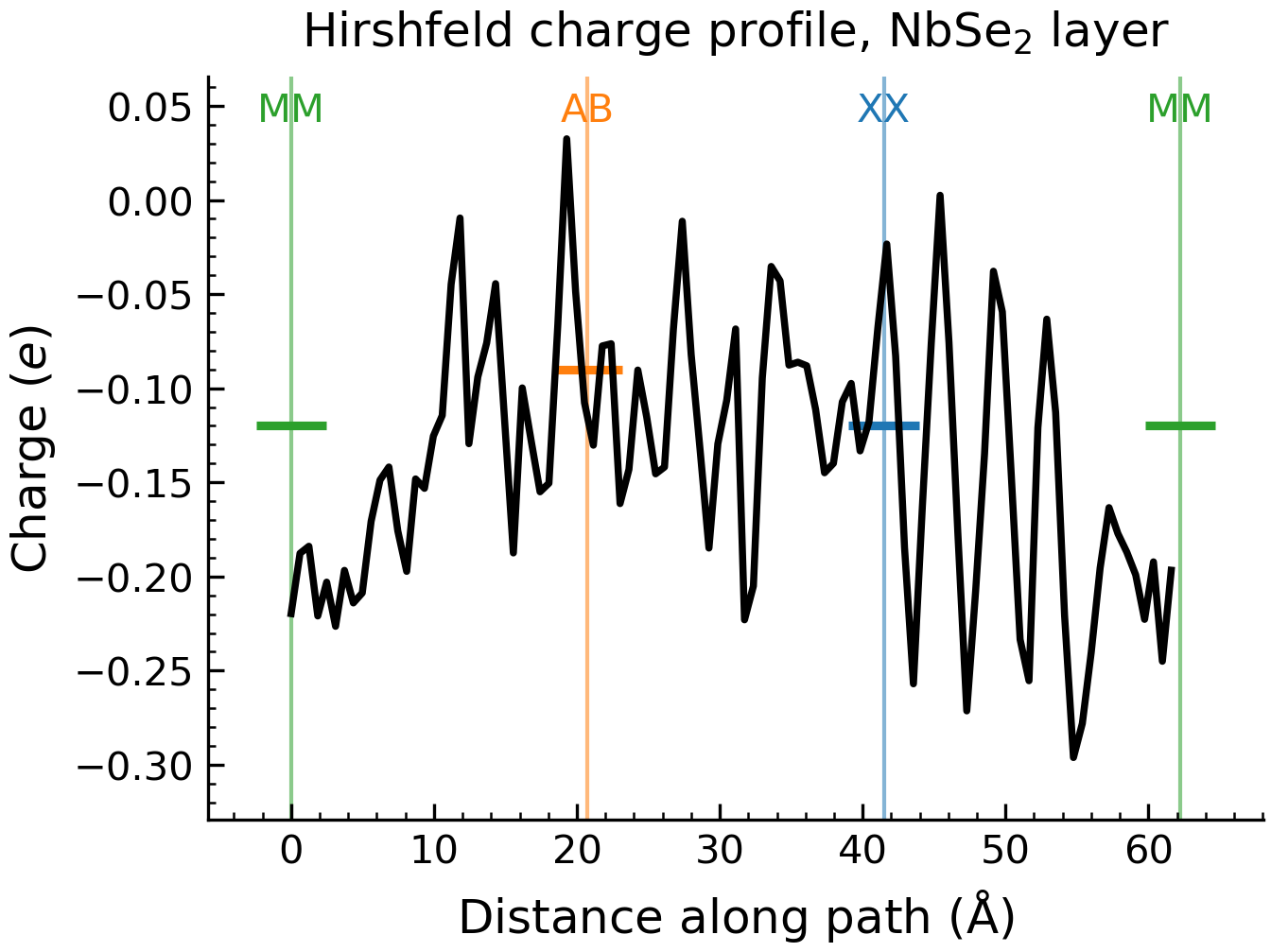}
    \end{subfigure}%
    \par
    \begin{subfigure}[h]{\figfacinplct\textwidth}
        \centering
        \caption{\hspace{\hsinplct} }
        \includegraphics[width=\textwidth]{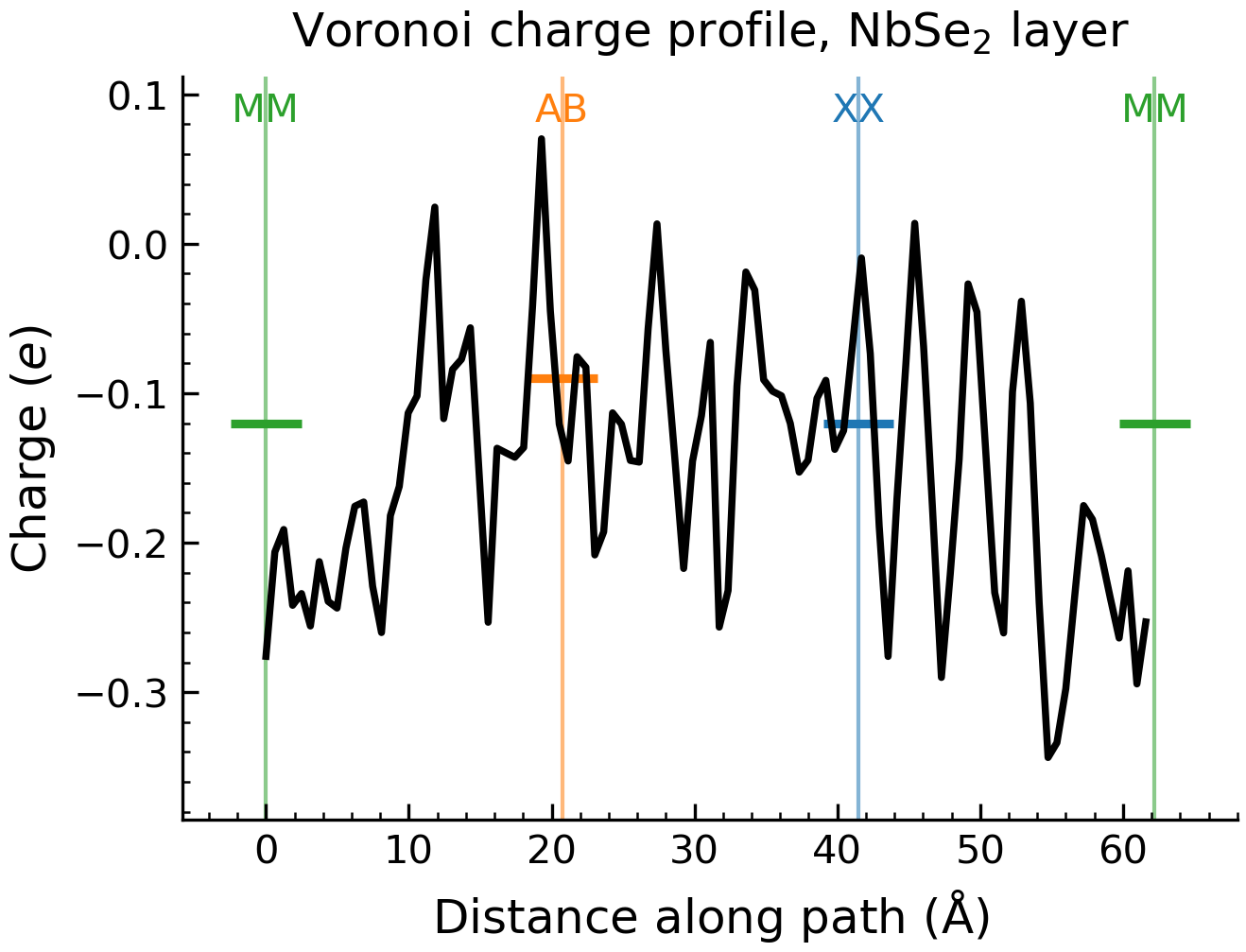}
    \end{subfigure}%
    \par
    \begin{subfigure}[h]{\figfacinplct\textwidth}
        \centering
        \caption{\hspace{\hsinplct} }
        \includegraphics[width=\textwidth]{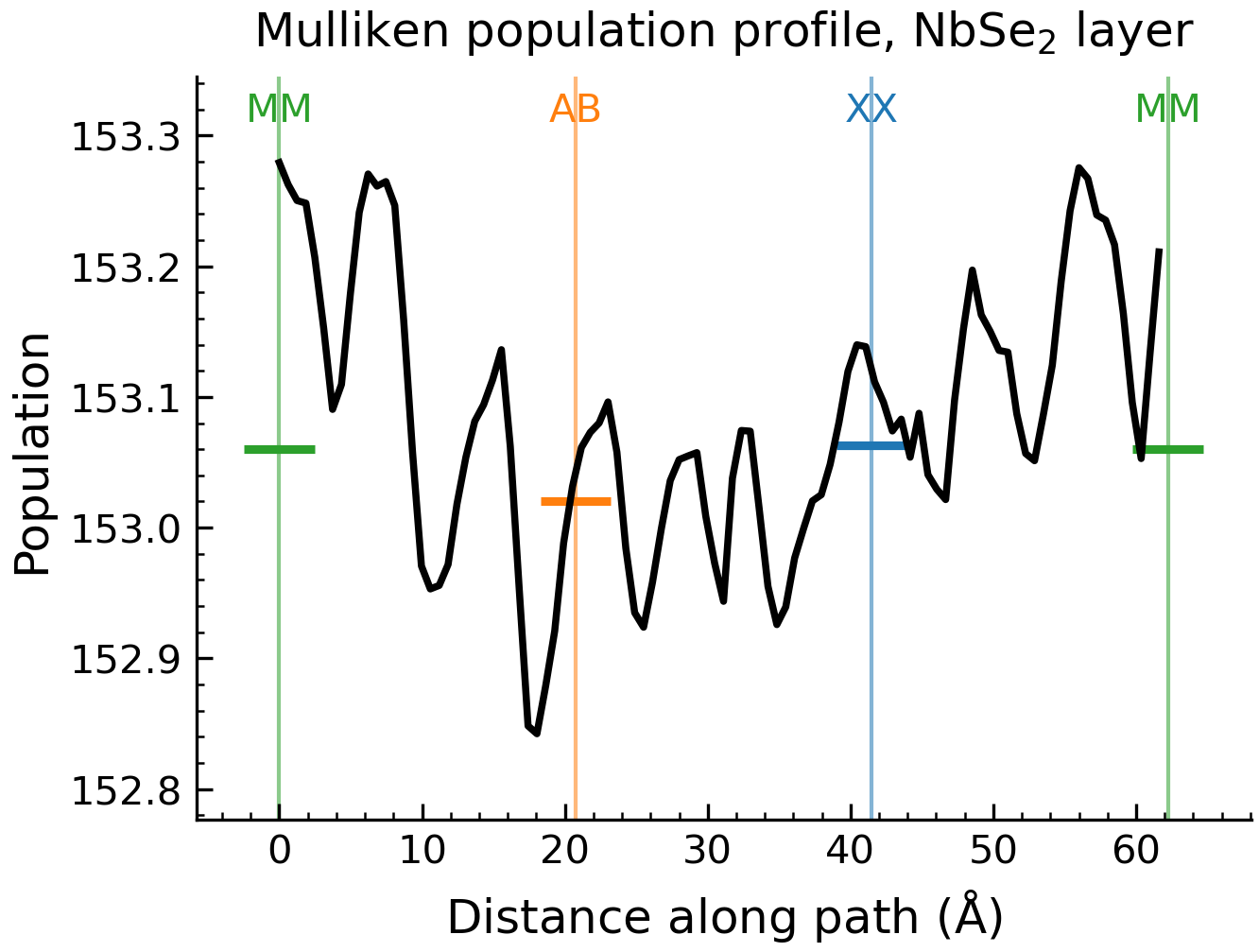}
    \end{subfigure}%
    \caption{The spatial variation of the charge per unit cell in the NbSe$_2$ layer along the diagonal of the moir\'e unit cell calculated using (a): Hirshfeld, (b): Voronoi, and (c): Mulliken charge partition schemes. The vertical lines indicate the high-symmetry centers, and the horizontal lines indicate the values of the local charges in the corresponding high-symmetry stacked untwisted bilayer.}
    \label{fig:q_inplane}
\end{figure*}

\clearpage
\section{Charge transfer estimation in untwisted bilayers}
\label{sec:ct_utb}
We calculate the charge transfer between the NbSe$_2$ layer and the MoSe$_2$ layer using the Hirshfeld, Voronoi, and Mulliken partition schemes implemented in SIESTA. We sum over the atomic charges in the NbSe$_2$ layer or MoSe$_2$ layer respectively. The total charges on each layer are the same, as expected from charge neutrality of the system. We report the dependence of the total charge in the MoSe$_2$ layer against the displacement amplitude of the CDW in the NbSe$_2$ layer in Supplementary Fig.~\ref{fig:amp_ct}. 

We find that amount of interlayer charge transfer is estimated to be around $0.1~e$ per $3\times3$ supercell using the Hirshfeld and Voronoi schemes. The amount of charge transfer is also independent of the CDW amplitude in the NbSe$_2$ layer. We then relaxed monolayer MoSe$_2$ doped to this level, and found that no CDWs are stabilized. The amount of interlayer charge transfer is therefore insufficient to induce the CDWs observed in the MoSe$_2$ layer. 

\newcommand{\figfacampct}{0.44}
\newcommand{\hsct}{54pt}
\begin{figure*}[htbp!]
    \centering
    \begin{subfigure}[h]{\figfacampct\textwidth}
        \centering
        \caption{\hspace{\hsct} Hirshfeld charge partition}
        \includegraphics[width=\textwidth]{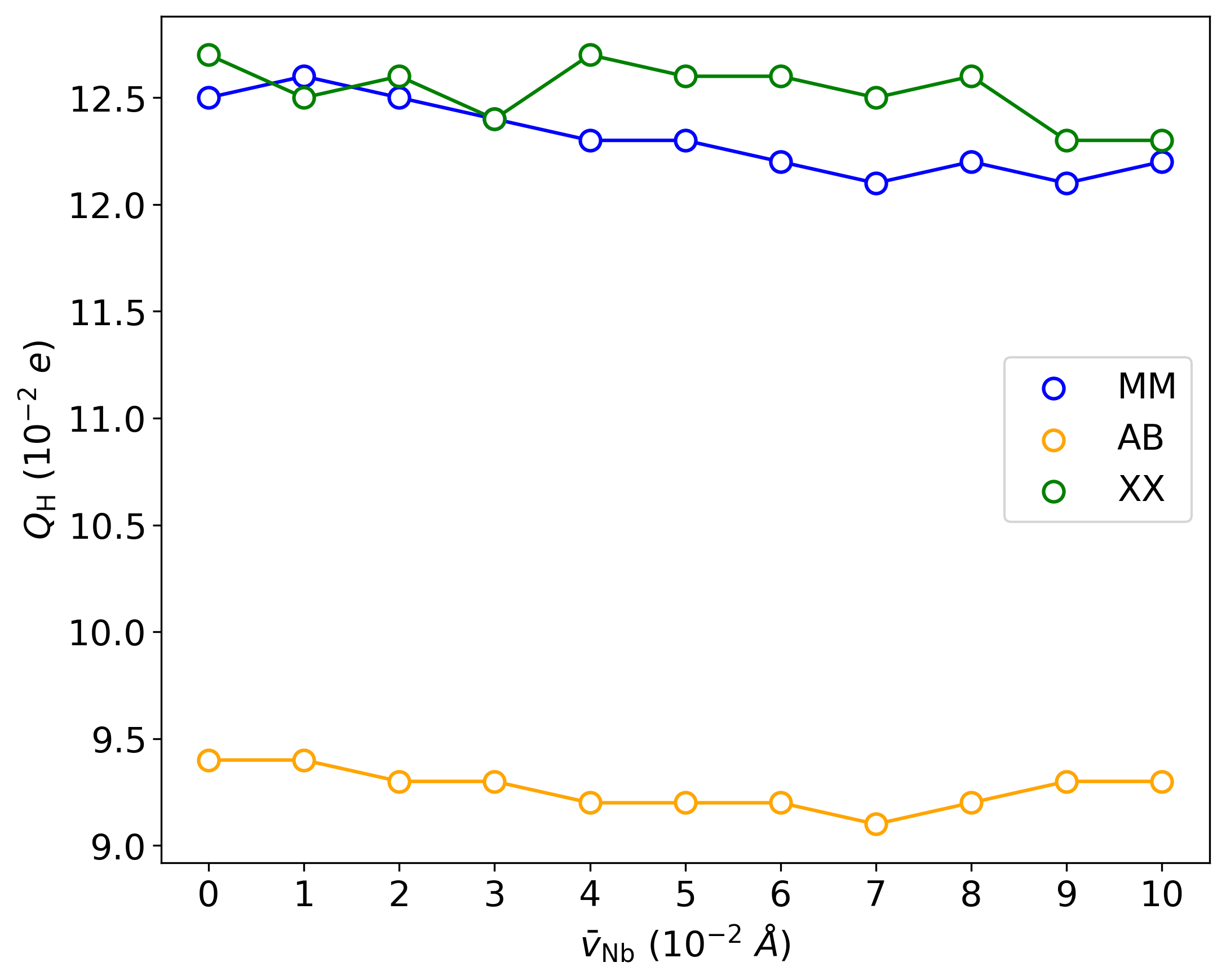}
    \end{subfigure}%
    \par
    \begin{subfigure}[h]{\figfacampct\textwidth}
        \centering
        \caption{\hspace{\hsct} Voronoi charge partition}
        \includegraphics[width=\textwidth]{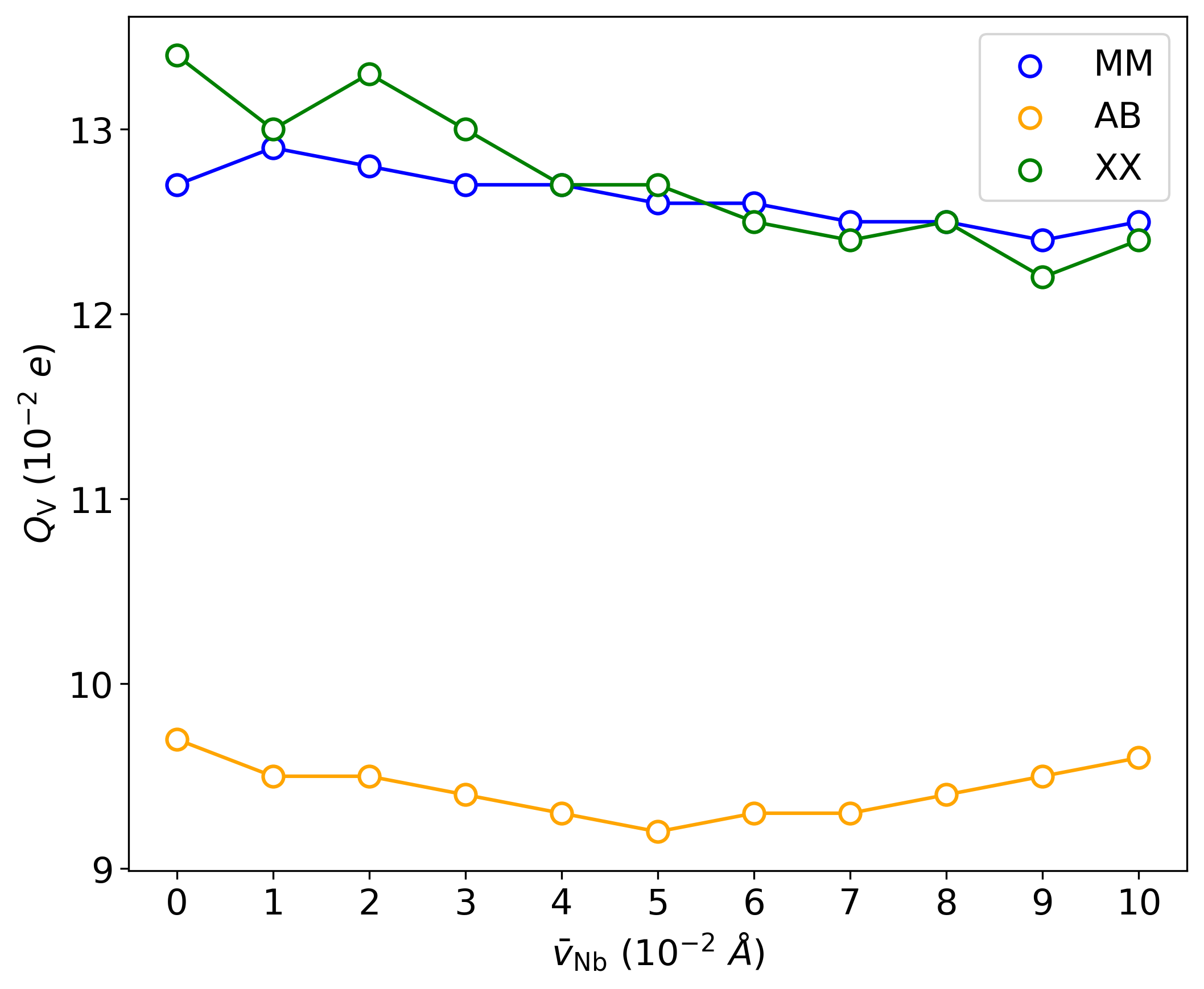}
    \end{subfigure}%
    \par
    \begin{subfigure}[h]{\figfacampct\textwidth}
        \centering
        \caption{\hspace{\hsct} Mulliken charge partition}
        \includegraphics[width=\textwidth]{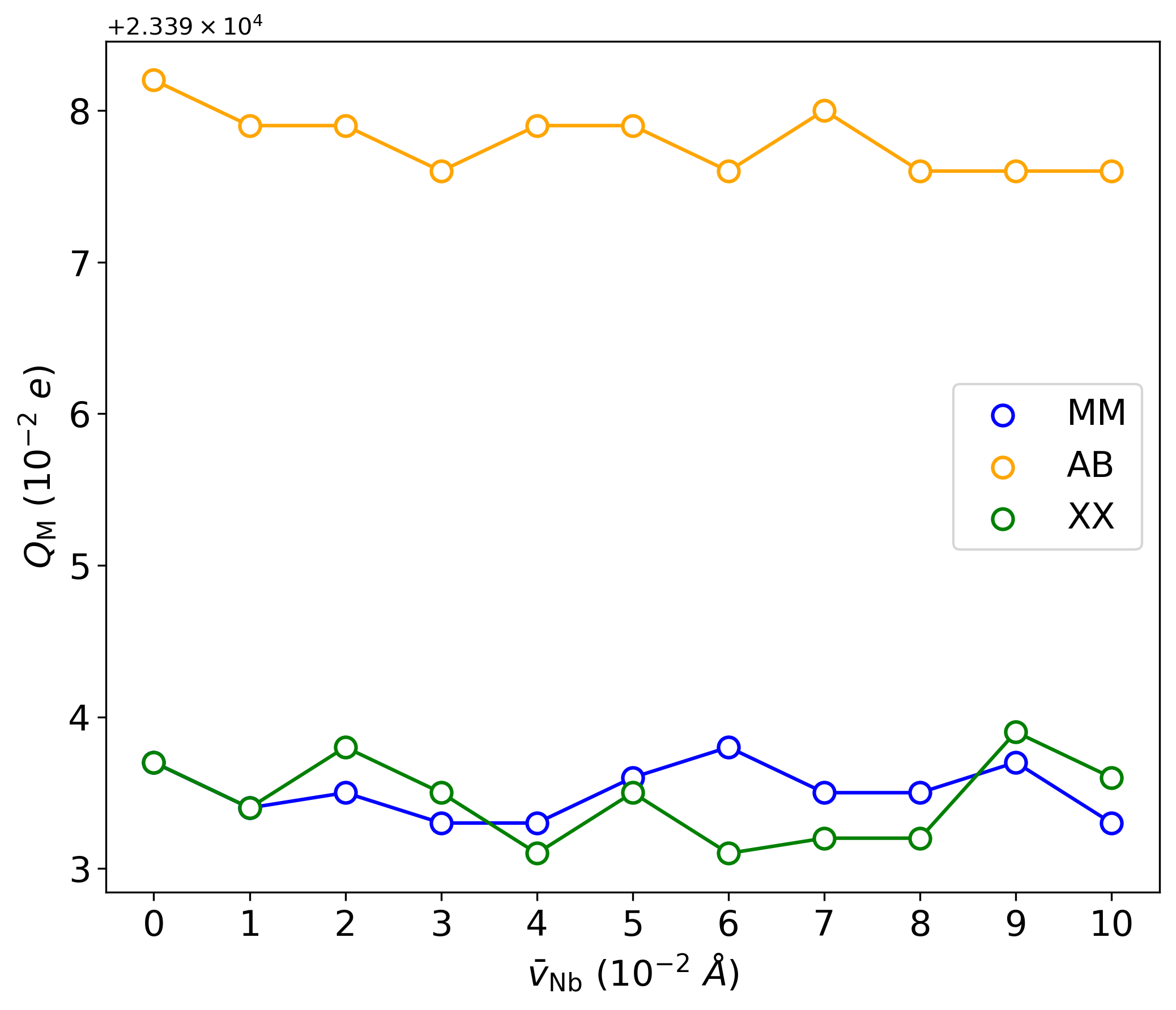}
    \end{subfigure}%
    \caption{The dependence of the total charge in the MoSe$_2$ layer calculated using (a): Hirshfeld, (b): Voronoi, and (c): Mulliken charge partition schemes against the CDW displacement amplitude in the NbSe$_2$ layer. }
    \label{fig:amp_ct}
\end{figure*}

\clearpage
\bibliography{cdw,dft,general_moire,optical_moire,vib_moire}

\end{document}

%% file: input.tex
\usepackage{amsmath,amsfonts,amsthm,bm,bigints,amssymb}
\newcommand{\vc}[1]{\mathbf{#1}}
\newcommand{\Ang}{\text{\AA}}

\newcommand{\Lagr}{\mathcal{L}}
\newcommand{\pos}{\mathbf{r}}
\newcommand{\atpos}{\mathbf{R}}
\newcommand{\den}{\rho}
\newcommand{\smearing}{\sigma}
\newcommand{\fitden}{\tilde{\rho}}
\newcommand{\phase}{\phi}

\newcommand{\penal}{\Delta}
\newcommand{\wavevec}{\mathbf{q}}
\newcommand{\normvec}{\mathbf{n}}
\newcommand{\wwavevec}{q}
\newcommand{\nnormvec}{n}

\newcommand{\lattconst}{a_0}
\newcommand{\ihat}{\hat{\textbf{\i}}}
\newcommand{\jhat}{\hat{\textbf{\j}}}
\newcommand{\denexp}{\exp\left(-\frac{(\vc{\pos}-\vc{\atpos}_i)^2}{2\smearing^2}\right)}

\newcommand{\displamp}{v}
\newcommand{\displphase}{\varphi}